\documentclass[a4paper,fleqn,usenatbib]{mnras}

\usepackage{newtxtext,newtxmath}

\usepackage[T1]{fontenc}
\usepackage{ae,aecompl}

\usepackage{graphicx}	
\usepackage{amsmath}	
\usepackage{amssymb}	

 \usepackage{ulem}

\title[A physical model of AGN SED]{A physical model of the broadband continuum
  of AGN and its implications for the UV/X relation and optical variability}

\author[A. Kubota \& C. Done.]{
Aya Kubota$^{1,2}$\thanks{E-mail: aya@shibaura-it.ac.jp }
and Chris Done$^{1}$
\\
$^{1}$Department of Physics, University of Durham, South Road, Durham, DH1 3LE,UK\\
$^{2}$Department of Electronic Information Systems, Shibaura Institute of Technology, 307 Fukasaku, Minuma-ku,   Saitama-shi, Saitama 337-8570, Japan\\
}

\date{Accepted XXX. Received YYY; in original form ZZZ} 

\pubyear{2018}

\begin{document}
\label{firstpage}
\pagerange{\pageref{firstpage}--\pageref{lastpage}}
\maketitle

\begin{abstract}

 We develop a new spectral model for the broadband spectral energy distribution (SED) of Active Galactic Nuclei (AGN). This includes an outer standard disc, an inner warm Comptonising region to produce the soft X-ray excess and a hot corona. We tie these together energetically by assuming Novikov-Thorne emissivity, and use this to define a size scale for the hard X-ray corona as equal to the radius where the remaining accretion energy down to the black hole can power the observed X-ray emission. 
 We test this on three AGN with well defined SEDs as well as on larger samples to show that the average hard X-ray luminosity is always approximately a few percent of the Eddington luminosity across a large range of Eddington ratio.
 As a consequence, the radial size scale required for gravity to power the X-ray corona has to decrease with increasing Eddington fraction.
 For the first time we hardwire this into the spectral models, and set the hard X-ray spectral index self consistently from the ratio of the hard X-ray luminosity to intercepted seed photon luminosity from the disc. This matches the observed correlation of steeper spectral index with increasing Eddington ratio, as well as reproducing the observed tight UV/X relation of quasars. We also include the reprocessed emission produced by the hot inner flow illuminating the warm Comptonisation and standard disc regions and show that this predicts a decreasing amount of optical variability with increasing Eddington ratio as observed, though additional processes may also be required to explain the observed optical variability.

\end{abstract}




\section{Introduction}

Active Galactic Nuclei (AGN) are powered by mass accreting onto a
supermassive black hole (SMBH). The well known \cite{ss73} disc model makes very simple predictions for
this emission if it is emitted locally and thermalises to a blackbody.
The disc temperature increases inwards (modulo a stress-free inner
boundary condition at the innermost stable circular orbit, $R_{\rm
  ISCO}$), so the total spectrum is the sum over all radii of these
different temperature components (multi-colour disc blackbody; e.g., \citealt{mitsuda1984}).
However, the observed spectral energy distribution (SED) of AGN are
much more complex than this predicts.  There is a ubiquitous tail at
X-ray energies, as well as an unexpected upturn below 1~keV, termed
the 'soft X-ray excess'.

The hard X-ray tail indicates that some part of the accretion energy is
not dissipated in the optically thick disc (where it would thermalise)
but is instead released in an optically thin region (e.g., \citealt{elvis1994}). The resulting Comptonised spectrum from 1--100~keV indicates
that this region has electron temperature $kT_e\sim 40$--100~keV and
optical depth $\tau\sim 1$--2
(\citealt{lubinski2016,fabian2015}). 

The origin of the 'soft X-ray excess' is not well understood. It can
be fit by a second Comptonisation region with very different
parameters from the coronal emission, one where the electrons are
warm, $kT_e\sim 0.1$--1~keV and optically thick $\tau\sim 10$--25
(e.g., \citealt{magdziarz1998,czerny2003,marek2004b,porquet2004,petrucci2013,middei2018}).
Alternatively, it could be produced by reprocessing/reflection
of the coronal emission on the very inner disc, where extremely strong
relativistic effects smear out the expected strong line emission from
ionised material \citep{crummy2006}.  
The fastest soft X-ray variability is correlated with, and lags
behind, the hard X-ray variability, so 
some fraction of the soft X-ray excess 
must be produced from reprocessing/reflection of the corona flux
(e.g., \citealt{fabian2013,demarco2013}). However, 
recent results have shown that the majority of the soft excess does
not arise from reflection
\citep{509,5548,noda2013,matt2014,boissay2016,porquet2018}, favouring
the warm Comptonisation model.

The warm Comptonisation scenario also helps to explain another puzzling
component of the broadband AGN SED, namely a ubiquitous downturn seen
in the UV, at energies far below those expected for the peak disc
temperature (e.g., \citealt{zheng1997,davis2007}).  A warm
Comptonisation spectrum can extend across the absorption gap,
connecting the UV downturn and the soft X-ray upturn with a single
component \citep{elvis1994,laor1997,richards2006}.
This carries a dominant fraction of the luminosity in the
SED of AGN at lower Eddington ratio, $L_{\rm bol}/L_{\rm Edd}$
\citep{jin2012a,jin2012b}, 
again arguing against a purely reprocessing/reflection origin for the soft X-ray excess, though
some contribution could be present 
(e.g., \citealt{lawrence2012}). The SED of high $L_{\rm bol}/L_{\rm Edd}$
are instead dominated by disk emission, which can extend into the soft
X-ray bandpass for the lowest mass, Narrow Line Seyfert-1 (NLS1) galaxies, but
these still have a small fraction of their bolometric power emitted in
a soft X-ray excess component
\citep{jin2012a, jin2012b,done2012,jin2013,matzeu2017}.

Neither of the Comptonisation components are well understood. 
However the warm Comptonisation region is especially problematic as, unlike the hot 
corona, it does not have a clear counterpart in the much lower mass
black hole binary systems (BHB). These often show spectra at $L_{\rm bol}/L_{\rm
  Edd}\sim 0.1$--0.2 which are dominated by the thermal accretion disc
emission, with only a small tail to higher energies from a hot
Comptonisating corona (e.g., \citealt{kubota2001,marek2004a,steiner2009}).

One obvious break in scaling between BHB and AGN is that the
SMBHs have discs which peak in the UV rather than
X-ray temperature range. The UV is a region in which atomic
physics is extremely important whereas plasma physics dominates in
BHB. Nontheless, the best models of the accretion disc structure
including UV opacities
\citep{hubeny2001} find that the spectra
are fairly well described by a sum of modified blackbody components
(with atomic features superimposed), similar to BHB
spectra \citep{davis2006}. The addition of UV opacity within the
disc alone then may not be enough to explain the soft X-ray excess
(though it does also depend on the heating profile within the disc):
instead it may be connected to the ability of UV line opacity to launch winds
from AGN discs  (e.g., \citealt{proga2000,laor2011}) and/or the huge change in
opacity connected to Hydrogen ionisation which may be able to change
the entire disc structure away from steady state models 
\citep{hameury2009}.

Constraining the shape of the warm comptonisation component is not
easy as it spans the 0.01--1~keV range where interstellar absorption
from our own Galaxy obscures our view.  Spectral fitting becomes
especially degenerate when trying to simultaneously constrain this
component along with the hotter coronal component and any residual
emission from an outer standard disc (e.g.,
\citealt{jin2009}). Instead, \cite{done2012} (hereafter D12) assumed
that the emission is ultimately powered by energy release from
gravity, with the same form as for the thin disc, but that the
dissipation mechanism is only blackbody for radii $R>R_{\rm corona}$,
Inwards of
this, they assumed that the flow instead emits the accretion energy as
a warm or hot Comptonisation component.  These energy conserving
models ({\sc optxagnf}: D12) give an additional physical constraint on
the components, and more importantly, highlight the fundamental
parameters of mass and mass accretion rate (for any assumed spin) in
setting the overall SED
(\citealt{jin2012a,jin2012b,ezhikode2017}). These models reveal a
systematic change in the SED which can be modeled by a decrease in
$R_{\rm corona}/R_{\rm g}$  (where $R_{\rm
    g}=GM/c^2$) correlated with a increase in the hot Comptonisation
power law spectral index as $L_{\rm bol}/L_{\rm Edd}$ increases
(\citealt{jin2012a,jin2012b,ezhikode2017}; see also
\citealt{shemmer2006,shemmer2008} and
\citealt{vasudevan2007,vasudevan2009} for the hard X-ray spectral
index).

In this paper, we develop a new model which addresses the underlying physics of
these changes, where we assume that the flow is completely radially
stratified, emitting as a standard disc blackbody from $R_{\rm out}$ to 
$R_{\rm warm}$, as warm Comptonisation from $R_{\rm warm}$ to $R_{\rm hot}$
and then makes a transition to the hard X-ray emitting hot
Comptonisation component from $R_{\rm hot}$ to $R_{\rm ISCO}$.  
The warm Comptonisation component is optically thick, so we associate this with
material in the disc. Nonetheless, the energy does not 
thermalise to even a modified blackbody, perhaps indicating that
significant dissipation takes place within the vertical structure of
the disc, rather than being predominantly released in the midplane
(e.g., \citealt{davis2005}).
At a radius below $R_{\rm hot}$, 
the energy is emitted in the hot Comptonisation component. This
has much lower optical depth, so it is not the disc itself. It could
either be a corona above the inner disc, or the disc 
could truncate, so that the hot material fills the inner region close
to the black hole. We show that the observed steepening of the 2--10~keV
spectral index with increasing $L_{\rm bol}/L_{\rm Edd}$ can be most easily
explained with a true truncation. 

We describe the model structure in section 2, and apply it to observed
broadband spectra of individual AGN in section 3. We use these data to
set some of the model parameters, so that we can predict the entire
AGN SED as a function of only mass and mass accretion rate (for a
given black hole spin) in section 4. In section 5, 
we show that these models
reproduce the observed tight relationship between the UV and X-ray
emission in Quasars \citep{lusso2017}  as well as predict a
decrease in the fraction of reprocessed optical variability with
increasing $L_{\rm bol}/L_{\rm Edd}$ as observed.
Thus, this AGN SED model succeeds in describing multiple disparate
observational trends, which gives confidence that the 
assumed geometry captures most major aspects of the source
behaviour.

\section{Overall disc model}
\label{sec:overall}

We follow D12 and assume a radial emissivity like Novikov-Thorne
(hereafter NT), defining the flux per unit area at a radius $R$
on the disc as $F_{\rm NT}(R)=\sigma T_{\rm NT}^4(R)$, where
  $T_{\rm NT}(R)$ is the effective temperature at this radius. 
Converting to dimensionless units, with $r=R/R_g$,
$\dot{m}=\dot{M}/\dot{M}_{\rm Edd}$ and $L_{\rm Edd}=\eta\dot{M}_{\rm
  Edd}c^2$ gives $F_{\rm NT}\propto (\dot{m}/M)r^{-3}$ for $r>>6$.
Here $\eta$ is a spin dependent efficiency factor, assumed
fixed at 0.057 for a non-spinning black hole throughout
  this paper. 

\subsection{Standard Disc and warm Comptonisation region}

\begin{figure}
\begin{center}
	\includegraphics[width=0.95\columnwidth]{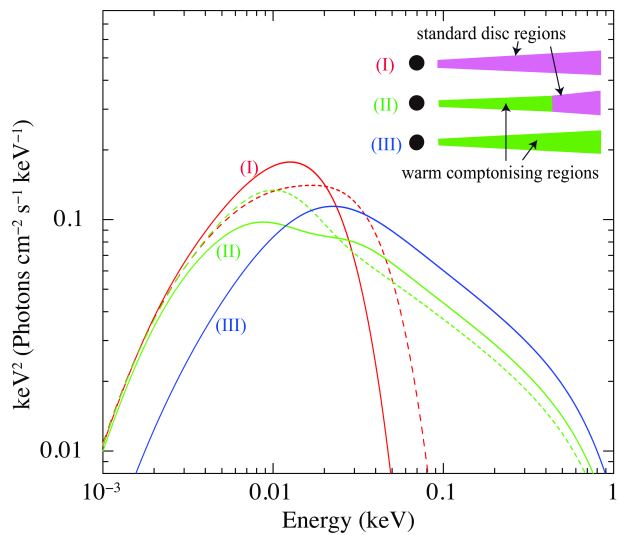}
\end{center}
\caption{Comparison of spectra for a black hole of $M= 10^8~M_\odot$
  with $\dot{m}=0.05$. Geometry I shows the Novikov-Thorne disc
  extending down to $R_{\rm ISCO}$ without (red solid) and with (red
  dashed) color-temperature correction. Geometry II shows an outer
  Novikov-Thorne disc plus a warm Comptonisation region from
  $r=40$ down to $r_{\rm ISCO}$. The green solid line shows
  {\sc agnsed} model used in this paper (see section 2.3) where seed photons are from the underlying cool
  material in the midplane compared to the {\sc optxagnf} assumption
  of seed photons from the inner edge of the standard disc (green
  dashed). Geometry III is complete coverage of the warm Compton region
  over the entire NT disc (blue), which has lower normalisation due to
the Compton scattering. We assume a photon index and electron temperature of the warm Comptonising corona of 2.5 and 0.2~keV, respectively.}    \label{fig:comparison}
\end{figure}

In the standard disc region we assume that the NT emission thermalises
locally either to give a blackbody $B_\nu(T_{\rm NT})$ at the local blackbody
temperature, defined from ${F}_{\rm NT}(R)=\sigma T_{\rm NT}^4(R)$, or
that electron scattering within the disc distorts this into a modified
disc blackbody spectrum. This can be approximated as a colour
temperature corrected blackbody, $B_\nu(f_{\rm col}T_{\rm NT})/f_{\rm col}^4$,
where $f_{\rm col}$ depends on the importance of electron scattering
compared to true absorption processes, which itself depends on disc
temperature, especially close to Hydrogen ionisation at $\sim
10^4$~K. There are few free electrons below this temperature, so
electron scattering is not important, and $f_{\rm col}\sim 1$, whereas
above this temperature there are multiple free electrons so $f_{\rm
  col}> 1$.  This effectively shifts the peak of the blackbody over by
a factor $f_{\rm col}$, and reduces its norm by a factor $f_{\rm col}^4$ so
this gives a shift to higher energy and decrease in normalisation in
the disc spectra from each annulus which onsets at around the hydrogen
ionisation energy. Thus the standard disc with this colour temperature
correction always has less UV emission (shortwards of $\sim
10^{15}$~Hz $\approx$ 2000\AA) than predicted from simple models with $f_{\rm
  col}=1$ (D12).  
Figure~\ref{fig:comparison} shows a comparison of the
standard disc (geometry I) with $f_{\rm col}=1$ (red solid) to that
where $f_{\rm col}(T_{\rm NT})$ is
derived from an analytic treatment of the vertical structure of the
disc (dashed red line, see also 
\citealt{davis2006}, D12). This clearly shows how the
outer disc emission is identical, while the inner disc emission is shifted to
higher temperatures/lower luminosities. 

Concerning the warm Comptonising region, the UV data do indeed show a downturn, but this is stronger than predicted by the effect of a changing
$f_{\rm col}$ in general. 
\cite{davis2007} show that the observed AGN spectra have
redder UV slopes than predicted from disc models even including
electron scattering. Instead, what is required to fit this UV downturn
is that the emission is much more strongly distorted from a blackbody
than predicted in the standard disc. While this could be modeled by a
larger colour temperature correction, a shifted blackbody becomes a
progressively poorer approximation for the spectrum as $f_{\rm col}$
increases. Hence we replace $f_{\rm col}$ with a fully Comptonised shape and do
not include this factor in our new code as the disc vertical structrue
is clearly very different to that of \cite{ss73}. 
Comptonisation also gives the possibility to connect the observed
downturn in the UV to an upturn seen in the soft X-ray spectra,
forming a single component spanning the unobservable
EUV range (D12, \citealt{509,5548}). 
This warm Comptonising emission
could be produced if some fraction of the dissipation takes place higher up
in the disc, rather than being concentrated towards the equatorial
plane \citep{czerny2003,rozanska2015}. Residual emission in the 
denser disc material on the midplane can then act as a 
source of seed photons, together with the reprocessed emission from
illumination from the upper layers of the disc \citep{petrucci2017}. 

Figure 14 in \cite{davis2005} shows the predicted (colour temperature
corrected) blackbody spectrum of a disc annulus where the vertical
dissipation goes with density as in standard disc models, compared to
one where the dissipation is arbitrarily changed so that 40\% of the
power is released in the photosphere (Fig. 16 of
\citealt{davis2005}). The spectrum is strongly Comptonised into a
steep tail to higher energies, but clearly contains the imprint of the
seed photons as a downturn at low energies. This seed photon
temperature is determined both by the intrinsic dissipation in the
lower layers of the disc (the remaining 60\% of the accretion power in
this specific example), and the thermalised flux resulting from
irradiation by the Comptonising upper layers. Both these physical
processes give seed photons which are close to the surface temperature
predicted by the standard disc dissipation, so the seed photon
temperature imprinted onto the steep Comptonised emission is itself
close to this temperature (Fig. 14 of \citealt{davis2005}).

Thus the expectation is that the seed photon energy should change with
radius in the same way as the expected standard disc temperature.  D12
discuss this in their Appendix, but make the simplifying assumption in
{\sc optxagnf} that this can be approximated as a single
Comptonisation spectrum with seed photon temperature set by the
maximum temperature of the standard disc emission
i.e. $kT_{\rm seed}=kT_{\rm NT}(R_{\rm corona})$. This is adequate if the low energy
part of this component is mostly unobservable due to interstellar
absorption. However, there are now data where this region of the
spectrum can be seen, motivating a more careful approach.  Also, the
{\sc optxagnf} approximation always requires that there is an outer
standard disc in order to provide the code with a temperature for the
seed photons. This need not be the case in the physical situation
envisaged. The warm Comptonisation region could instead cover the
entire outer disc as its seed photons are from deeper layers of the
underlying disc rather than from an external
source. 

\cite{petrucci2017} tested a model where the entire
optical/UV/soft X-ray flux is from a warm Comptonisation region with a slab geometry over the disc. 
They show that reprocessing in this geometry hardwires the Compton amplification
factor $A$ to 
\[
L_{\rm tot}=AL_{\rm seed} =L_{\rm seed} +L_{\rm diss,warm}
\] 
where, $L_{\rm tot}$, $L_{\rm seed}$ and $L_{\rm diss, warm}$ are
the total luminosity, seed photon luminosity underneath the Comptonising skin and power dissipated in the warm corona, respectively.
Their eq.(19) with the slab corona entirely covering the disc and large
optical depth with complete thermalisation, gives
\[
\frac{L_{\rm diss,warm}}{L_{\rm seed}}=A-1=2\left(1-\frac{L_{\rm diss, disc}}{L_{\rm seed,tot}}\right)-1
\] 
where $L_{\rm diss,disc}$ is the intrinsic dissipation which
thermalises in the disc. If there is no intrinsic
dissipation ('passive disc' on the midplane: \citealt{petrucci2017,petrucci2013})
then all these seed photons are set by thermalisation of the warm
Compton as $L_{\rm diss, disc}=0$.
Thus  
\[
\frac{L_{\rm diss,warm}}{L_{\rm seed}}=1
\]
hence $L_{\rm seed}=L_{\rm tot}/2$. This is emitted from the same surface area as the standard disc, so thus hardwires the seed photon temperature $T_{\rm seed}\simeq T_{\rm NT}$.


\cite{petrucci2017} showed that $A=2$ is equivalent to a photon index
of the warm compton component, $\Gamma_{\rm warm}=2.5$, which generally gave
a good fit to the observed soft X-ray excess component when combined
with a  Comptonising electron temperature
$kT_{e,{\rm warm}}\sim 0.2$~keV to give the observed 
rollover in soft X-rays.
Based on this passive disc picture, our new model, 
calculates the Compton emission at each annulus of radius $R$ and
width $\Delta R$
in the soft compton region, using the {\sc nthcomp} model \citep{zdziarski1996,zycki1999} in {\sc xspec}.
We set the seed photon temperature to the local disc temperature
$T_{\rm NT}(R)$, and set the local
luminosity to $\sigma T^4 _{\rm NT}(R)\cdot 2\pi R\Delta R\cdot 2$, and sum
over all annuli which produce the warm Comptonisation. 
Both photon index and electron temperature are free parameters, assumed to be the same for all the disc anulii which produce the warm Comptonisation.

The green solid line in Fig.\ref{fig:comparison} shows our new version
of the warm Comptonisation, summed over all radii from
$r_{\rm warm}=40 <r_{\rm out}$ to $r_{\rm ISCO}=6$ as in geometry II of
Fig.\ref{fig:comparison}.  We compare this to the  {\sc optxagnf} model
for the same parameters (green dashed line), showing the difference in
behaviour around the seed photon energy (see also Fig. A1b in D12).
The blue line in Fig.\ref{fig:comparison} corresponds instead to the
geometry of \cite{petrucci2017} sketched as geometry III in
Fig.\ref{fig:comparison}, i.e. where there is no outer disc so $r_{\rm
  warm}=r_{\rm out}$. The most noticeable effect is that the
normalisation of the SED in the optical/UV is 
reduced. This is important, as it changes the
otherwise quite robust relation between the luminosity in some band on
the low energy disc tail, and the mass accretion rate. The NT
emissivity sets the seed photon temperature and emissivity, but the
Comptonisation acts like a colour temperature correction and shifts
the entire spectrum to higher energies. 

We note here that unlike
\cite{petrucci2017}, our new model ties the seed photons for the warm
Comptonisation to the parameters of the underlying disc,
rather than allowing the seed photon temperature to be a free
parameter. 
Both the warm Comptonisation and standard disc are optically thick, so
we assume that the emission is  proportional to $\cos i$, where $i$ is the
inclination of the disc. 

\subsection{Hot Comptonisation region}
\label{sec:pl}

There is also an additional X-ray component which dominates over the
soft X-ray excess beyond 1--2~keV, extending up to
$kT_e\sim40$--$100$~keV, with $\tau\sim 1$--2
\citep{fabian2015,lubinski2016,petrucci2017}.  The low optical depth
clearly distinguishes this component from the disc material, and the
warm Compton region, so it needs to arise in a different
structure. This could either be a corona above a disc, with some
fraction of the accretion energy dissipated in this optically thin
material, or the optically thick disc could truncate, leaving a true
hole in the inner disc. For $\dot{m}\lesssim 0.2$, the hard X-ray
photon spectral index is usually $\lesssim 1.9$, i.e. it is flatter
than expected even in the limit where all the accretion energy is
emitted in the corona. Reprocessing and thermalisation of the (assumed
isotropic) illuminating flux even by a completely cold, passive disc,
sets a lower limit on $\Gamma_{\rm hot}\sim 1.9$
\citep{haardt1991,stern1995,malzac2005}.  Hence we assume that the
disc truly truncates at $r_{\rm hot}$ for low $\dot{m}$, as is
supported by the lack of strong reflection and lack of strong
relativistic smearing these AGN 
\citep{matt2014,yaqoob2016,porquet2018}.

In a truncated disc geometry, the seed photons seen by the hot flow
are predominantly from the inner edge of the warm Comptonisation
region, so these have typical seed photon energy of
$T_{\rm NT}(R_{\rm hot})\cdot \exp(y_{\rm warm})$, where 
$y_{\rm warm}=4\tau^2 kT_{e,\rm warm}/m_ec^2$ is the Compton 
$y$-parameter for the warm comptonisating corona.
 We use the {\sc xspec} model {\sc nthcomp} to
describe this, with total power
\begin{eqnarray}
\label{eq:Lhot}
L_{\rm hot}&=&L_{\rm diss,hot}+L_{\rm seed}
\end{eqnarray}
Here, the inner flow luminosity, $L_{\rm hot}$, is the sum of the
dissipated energy from the flow, $L_{\rm diss,hot}$, and the seed
photon luminosity which is intercepted by the flow, $L_{\rm seed}$.

\begin{figure}
\begin{center}
	\includegraphics[width=0.95\columnwidth]{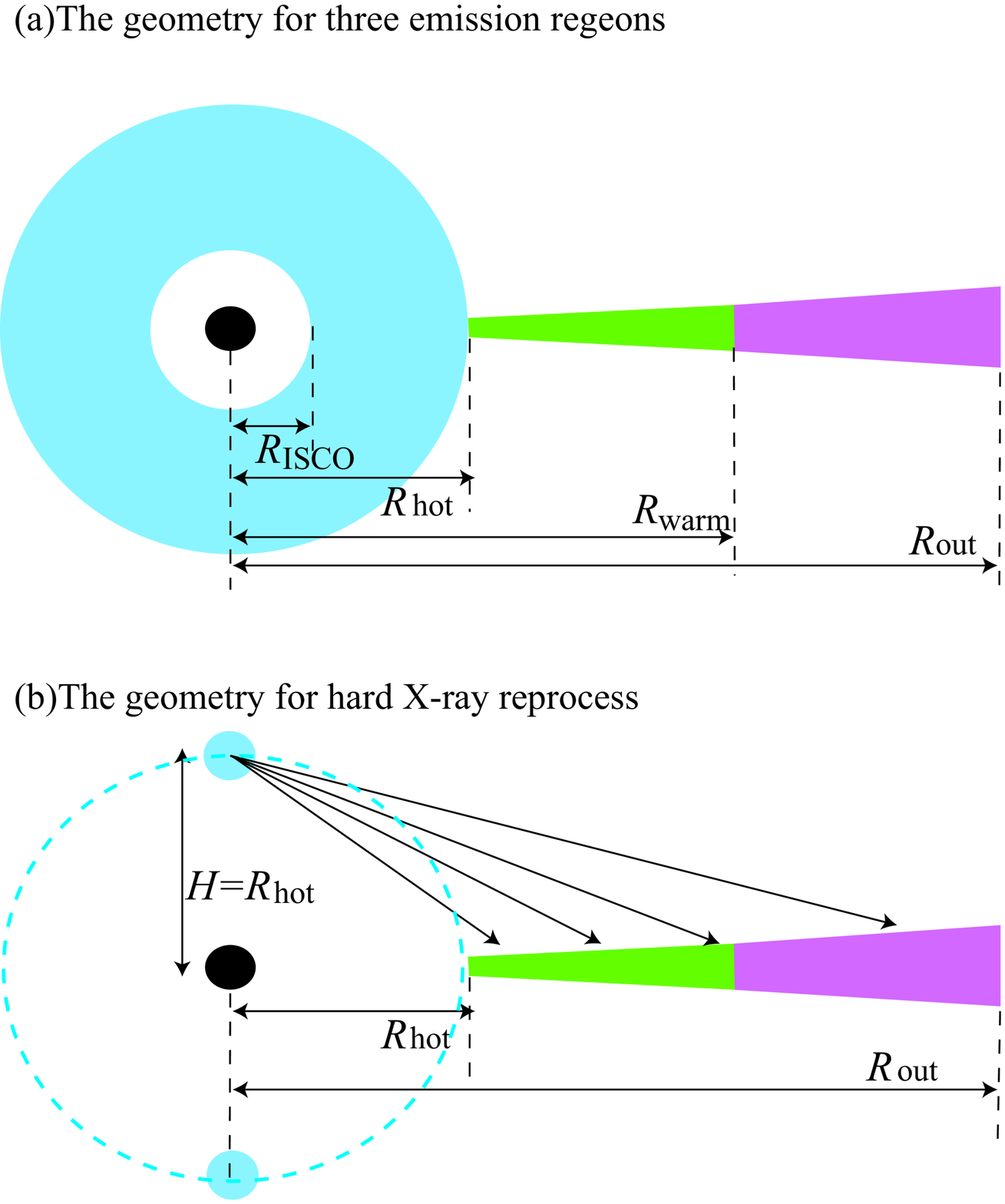}
\end{center}
    \caption{Geometry geometry of the model. (a) The geometry for hot
      inner flow (blue), warm Compton emission (green) and outer
      standard disc (magenta). (b) The lamppost geometry used to
      simplify the calculation of the reprocessed emission.}
    \label{fig:geometry}
\end{figure}

This gives $L_{\rm diss,hot}$ as
\begin{eqnarray}
\label{eq:Lhotdiss}
L _{\rm diss, hot}&=&2\int_{R_{\rm ISCO}}^{R_{\rm hot}} F_{\rm NT}(R)\cdot 2\pi R dR \nonumber \\
&=&2\int_{R_{\rm ISCO}}^{R_{\rm hot}} \sigma T_{\rm NT} (R)^4\cdot 2\pi R dR
\end{eqnarray}
with the  truncated radius $R_{\rm hot}$. This is shown geometrically
in Fig.\ref{fig:geometry}a. 
Since the X-ray emission is not very optically thick we assume it is
isotropic, unlike the disc/warm comptonisation region where we assumed a disc geometry.
$L_{\rm seed}$ is the intercepted soft luminosity from both 
the warm Comptonising region and the 
outer disk. This can be calculated assuming a
truncated disc/spherical hot flow geometry (Fig.\ref{fig:geometry}a, i.e. a flow scale height 
$H\sim R_{\rm hot}$) as
 \begin{eqnarray}
\label{eq:Lseed}
L_{\rm seed}&=&2\int_{R_{\rm hot}} ^{R_{\rm out}}  (F_{\rm NT}(R)+F_{\rm rep}(R))\frac{\Theta (R)}{\pi} 2\pi R dR\\
\Theta(R)&=&\theta_0-\frac{1}{2}\sin 2\theta_0
\label{eq:theta}
\end{eqnarray}
Here, $\Theta(R)/\pi$ is the covering fraction of hot flow as seen from
the disc at radius $R>R_{\rm hot}$ with $\sin \theta_0=H/R$,
and $F_{\rm NT}(R)+F_{\rm rep}(R)$ is the flux from the warm
Comptonised and/or outer disc including reprocessing (discussed in the 
following section). 

We caution that there can be many 
other factors which influence  $L_{\rm seed}$ e.g. overlap of the disc and hot flow
\citep{zdziarski1999} and/or any radial/vertical gradient in the structure of the hot flow. 
Nonetheless, we start from the simplest  possible assumption which is a spherical, homogeneous source
with $H=R_{\rm hot}$ but we leave $H$ as a parameter in the following 
equations so that it can be used as a tuning parameter for alternative geometries by 
changing the seed photons intercepted by the source. Smaller $H$ gives a smaller solid angle, so fewer
seed photons and harder spectral indices, but has no effect on $R_{\rm hot}$ as this is set by the energetics. 

\subsection{Modeling reprocessing }
\label{sec:reprocess}

The assumed geometry  shown in Fig.~\ref{fig:geometry}a has
some fraction of the hot Comptonisation illuminating the warm
Comptonisation and cool outer disc regions. We include this self
consistently, so that the irradiating flux increases the local flux
above that given by the intrinsic $\dot{M}$.  
Though our geometry assumes
that the hot corona is an extended source, with $H\sim R_{\rm hot}$
as above
(see Fig.~\ref{fig:geometry}a), \cite{gardner2017} show that
illumination from an extended source can be well approximated by 
a point source at height $H$ on the spin axis
(lampppost: Fig.~\ref{fig:geometry}b).  We thus utilized the lamppost
geometry for the hot inner flow to calculate the reprocessed flux as
it is simpler than integrating over an extended source.  

The reprocessed flux for a flat disc at a
radius $R$ is then written as
\begin{eqnarray}
\label{eq:frep}
F_{\rm rep}(R)&=&\frac{\frac{1}{2}L_{\rm hot}}{4\pi(R^2+H^2)} \frac{H}{\sqrt{R^2+H^2}}(1-a)\nonumber\\
&=&\frac{3GM\dot{M}}{8\pi R^3} \frac{2L_{\rm hot}}{\dot{M}c^2}\frac{H}{6R_{\rm g}}(1-a)\left[1+\left(\frac{H}{R}\right)^2\right]^{-3/2}  
\end{eqnarray}
where $a$ is the reflection albedo. 
Hence the local flux at radius 
$R(>R_{\rm hot})$ is $\sigma T_{\rm eff}(R) ^4=F_{\rm NT}(R)+F_{\rm
  rep}(R)$.
Both $F_{\rm NT}(R)$ and $F_{\rm rep}(R)$ depend on radius as $R^{-3}$
for $R\gg R_{\rm g}$, thus the effect of the reprocessing is basically
to increase the flux across both the standard and warm Comptonised
disc by the same factor which is of order $1 + H/6R_{\rm g}$. Hence a
larger X-ray source increases the fraction of X-ray power which
illuminates the disc, as well as increasing the fraction of bolometric
power which is dissipated in the X-ray region via the change in
$R_{\rm hot}=H$.

\begin{figure}
\begin{center}
	\includegraphics[angle=0,width=0.95\columnwidth]{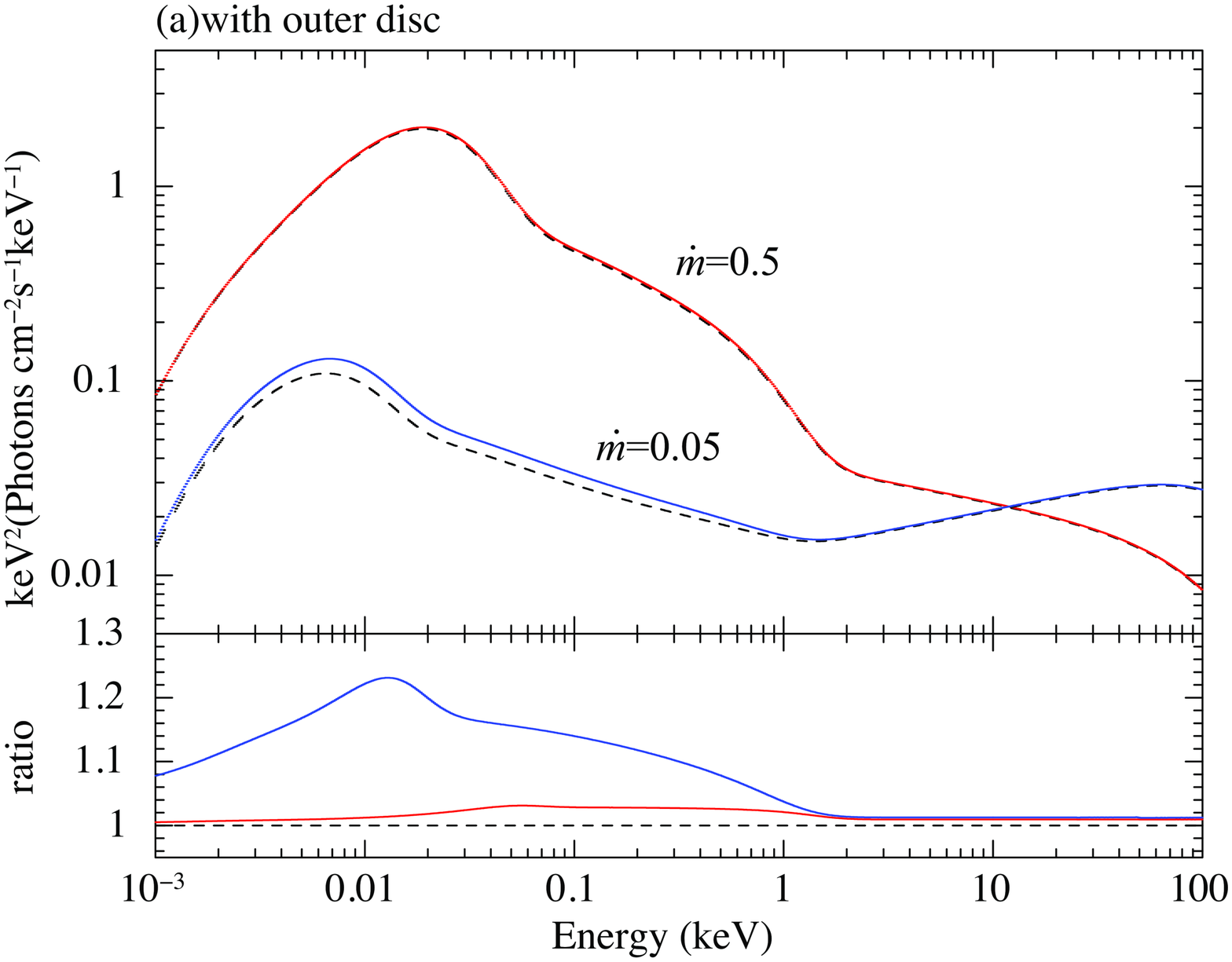}\\
	\includegraphics[angle=0,width=0.95\columnwidth]{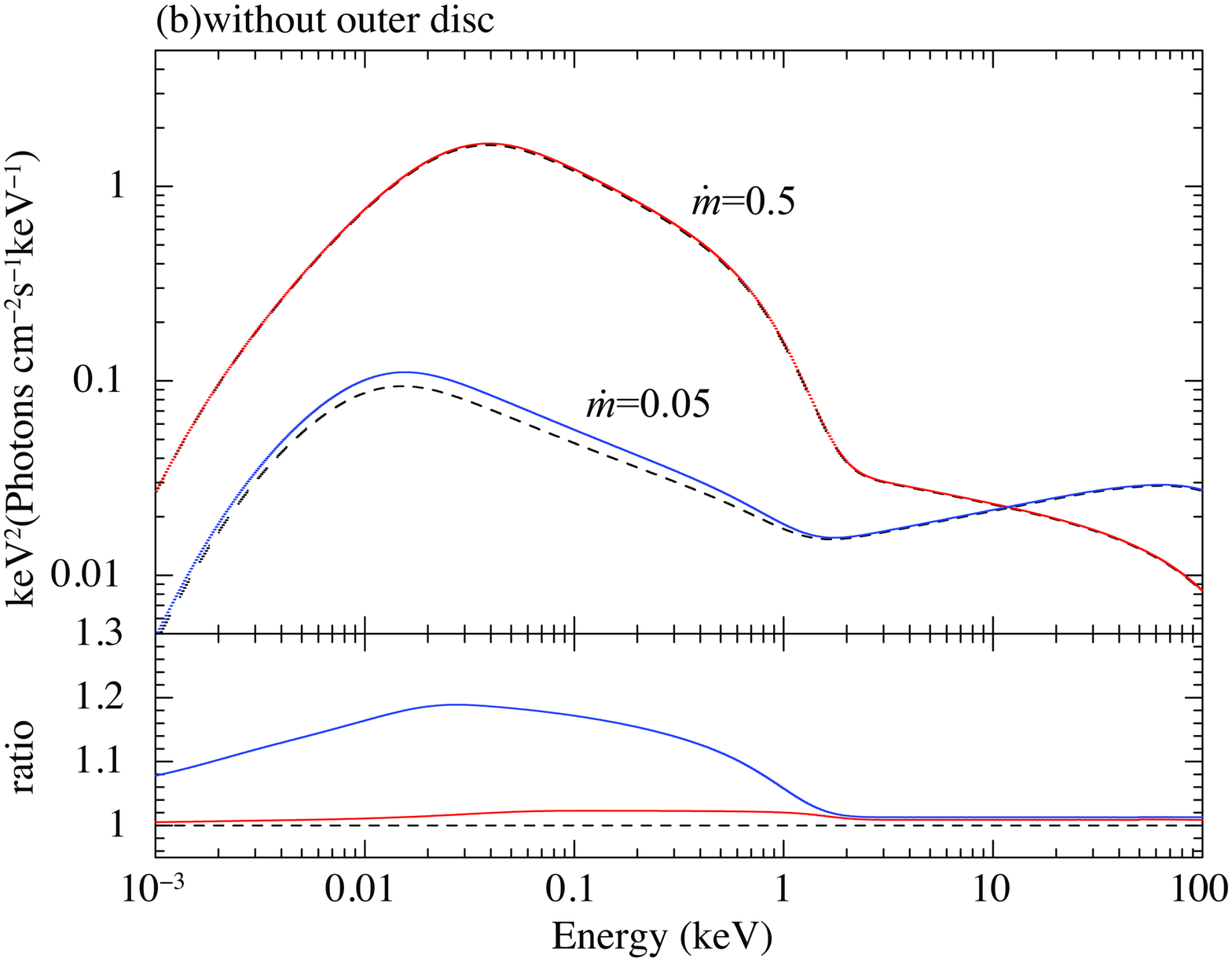}
\end{center}
\caption{comparison of spectra with
(solid) and without (dashed) reprocessing for a black hole of $M=10^8~M_\odot$ with $\dot{m}=0.05$
(blue) and 0.5 (red) at a distance 100~Mpc. 
The values of $kT_{e, {\rm warm}}$, $\Gamma_{\rm warm}$, $kT_{e, {\rm hot}}$ and 
$L_{\rm diss, hot}$ are assumed to be 0.2~keV, 2.5, 100~keV and $0.02L_{\rm Edd}$, respectively. The values of 
$\Gamma_{\rm hot}$ are assumed to be 1.8 and 2.2 for 
$\dot{m}=0.05$ and 0.5, respectively. 
Panel (a) and (b) correspond to with and without outer standard disc, respectively. 
In panel (a), $r_{\rm warm}$ is set to be $2r_{\rm hot}$.
Ratios of reprocessed
to intrinsic emission are also shown at each energy. }
\label{fig:reprocess}
\end{figure}

Figure~\ref{fig:reprocess} shows a comparison of example spectra with
(solid) and without (dashed) reprocessing for a black hole of
$10^8M_\odot$ with $\dot{m}=0.05$ (blue) and $0.5$ (red).  We set
$L_{\rm diss, hot}=0.02~L_{\rm Edd}$ and $kT_{e,\rm hot}=100$~keV for
both, which implies $r_{\rm hot}=23$ and $9$ for
$\dot{m}=0.05$ and $0.5$, respectively.  $\Gamma_{\rm hot}$ is set to
be 1.8 and 2.2 for $\dot{m}=0.05$ and $0.5$, respectively. We also
include a warm Comptonisation region with $kT_{e,{\rm warm}}=0.2$~keV and
$\Gamma_{\rm warm}=2.5$. Figure~\ref{fig:reprocess}a shows models
where the warm Comptonisation region extends from $r_{\rm hot}$ to
$r_{\rm warm}=2r_{\rm hot}$, so that there is a standard outer disc
region from $r_{\rm warm}$ to $r_{\rm out}$, while
Fig.~\ref{fig:reprocess}b shows the alternative geometry where the
warm Comptonisation extends over the entire outer disc, i.e. $r_{\rm
  warm}=r_{\rm out}$.

The lower panels of Fig.~\ref{fig:reprocess}a and b highlight the effect of
reprocessing by showing the ratio of the spectra including
reprocessing to the intrinsic emission.  Reprocessing makes a 
larger fraction of the optical emission for the lower $\dot{m}$, as
here the ratio of the X-ray flux to optical disc emission is much
larger, and the larger size scale of the X-ray source means that a
larger fraction of the X-ray emission is intercepted by the disc. The
difference is most marked around the maximum in the SED as the flux
increase from reprocessing is enhanced by the associated temperature
increase in disc or seed photon energy.

We call this new model {\sc agnsed}, and in the appendix define all
the parameters. 
The model is publicly available for use in the 
{\sc xspec} spectral fitting package. 
We also release a simplified model {\sc qsosed} where many of the parameters are fixed. 
This is more suitable for fitting fainter objects such as distant quasars, where the signal to noise is limited.

\section{Application to observed spectra}
\label{sec:application}

We apply {\sc agnsed}   to a small sample of AGN, spanning a wide range of $\dot{m}$,
 chosen to have  good multi-wavelength data.
We select NGC~5548 ($\dot{m}\sim0.03$; \citealt{5548}) and Mrk~509 ($\dot{m}\sim0.1$; \citealt{509}), for which 
big multiwavelength observation campaigns have been performed. 
Based on the long-term observations with the Reflection Grating Spectrometers (RGS) on board XMM-Newton, their intrinsic absorption were  
extremely well determined \citep{5548,detmers2011,509} and removed from the SEDs.
In this paper, we fit the best estimate of
the continuum spectra from these AGN, deconvolved from the instrument
response, and corrected for reddening and absorption. We read the
resulting flux files into {\sc xspec} using the {\sc flx2xsp} command.
These deconvolved data were kindly provided by M. Mehdipour.
In order to apply the model to  
higher $\dot{m}$ AGN, we select PG~$1115+407$ from 51 AGN sample analyzed by 
\cite{jin2012a}.
This object has little intrinsic absorption with $\dot{m}\sim 0.4$ \citep{jin2012a}, and emission from the host galaxy 
is negligible in band of  the optical monitor (OM) onboard XMM-Newton \citep{ezhikode2017}.


The system parameters for each AGN are given in table
\ref{tab:system}.  We fit all three SED with {\sc agnsed}   with
$i=45^\circ$ and limit the mass to the uncertainty range given in 
 table \ref{tab:system}. The outer disc
radius, $r_{\rm out}$, is initially set to equal to the self-gravity $r_{\rm sg}$ \citep{laor1989}.

\begin{table}
	\centering
	\caption{System parameters to calculate each spectrum. Comoving radial distance $D$ and 
	luminosity distance $D_{\rm L}$ is calculated based on $H_0=69.6~{\rm km~s^{-1}Mpc^{-1}}$, 
and $\Omega_{\rm M}=0.286$ with flat universe \citep{wright2006}. 
}
	\label{tab:system}
	\begin{tabular}{llccc} 
		\hline
& &NGC 5548&Mrk 509&PG~$1115+407$\\ 
\hline \hline
$z$ &&0.017175&0.034397&0.154338\\
$D$&Mpc&73.7&147.1&642.1\\ 
$D_{\rm L}$&Mpc&75.0&152.1&741.2 \\ 
\hline
$M$&$10^7M_\odot$&2--6& 10-30 &4.6--14\\
\hline
\multicolumn{2}{l}{Eddington ratio}&$\sim0.03$&$\sim0.1$&$\sim0.4$\\
\hline
\multicolumn{2}{l}{observation} &2013&2009&2002\\
\multicolumn{2}{l}{reference} &(1), (2)&(3), (4)&(5), (6)\\
\hline
\multicolumn{5}{l}{$^{(1)}$\citealt{kaastra2014}$^{(2)}$\citealt{5548} $^{(3)}$\citealt{kaastra2011-1} 
}\\
\multicolumn{5}{l}{
$^{(4)}$\citealt{509}  
$^{(5)}$\citealt{jin2012a}$^{(6)}$\citealt{jin2012b}}
	\end{tabular}
\end{table}

\subsection{NGC~5548: $\dot{m}\sim0.03$}
\label{sec:ngc5548}


The Seyfert-1 galaxy NGC~5548 is one of the most widely studied nearby
AGN, with well constrained mass (2--6)$\times 10^7M_\odot$.  There was
a multi-wavelength campaign on this object in 2012--2014 (e.g.,
\citealt{kaastra2014}), and the broadband spectra were analyzed by
\cite{5548}.
We use the data from summer 2013 shown in Fig.10 of
\cite{5548},
which includes data points from  NuSTAR, INTEGRAL,  RGS and the European Photon Imaging Camera 
(EPIC-pn) on XMM-Newton, the Cosmic Origins Spectrograph (COS) on the Hubble Space Telescope (HST) , 
the UltraViolet and Optical Telescope (UVOT) on  Swift, and from two ground-based optical observatories: the Wise Observatory (WO) and the Observatorio Cerro Armazones (OCA). 
During this campaign, there are multiple absorption systems seen in the X-ray  \citep{5548,kaastra2014,cappi2016}, 
which have been removed from our data using the best fit modelling of the RGS spectra
\citep{5548}. Nonetheless, 
there is some uncertainty associated with this, which affects the 
determination of the intrinsic soft X-ray spectrum.

There is also strong
host galaxy contamination in the optical, so this was removed
(\citealt[Fig. 10]{5548}) to isolate the AGN emission.  The resulting
optical/UV continuum is rather blue, and cannot easily be fit with any
disc blackbody based model, either a standard disc or warm Comptonised
one.  \cite{5548} fit this by a single, warm Comptonisation region,
using blackbody (not disc blackbody) seed photons in order to get such
a steeply rising optical/UV continuum. It seems more likely that this
is a consequence of a slight oversubtraction of the host galaxy, so we
ignore the V and I band continuum points in the fit. The X-ray
emission is extremely bright compared to the optical/UV, and very
hard.

We then fit the data with {\sc agnsed}, including three emission
regions. There is a clear iron line in the X-ray data, but the accompanying 
Compton hump is rather weak \citep{ursini2015,cappi2016} so this line probably originates in
optically thin material in the BLR \citep{yaqoob2001,brenneman2012,ursini2015}. 
We model this simply by
including a gaussian in the fit.  The model overpredicts the optical
data for $r_{\rm out}=r_{\rm sg}=880$ for $\dot{m}\sim
0.03$. This could again indicate a slight oversubtraction of the host
galaxy from our data, but we are able to fit by allowing the outer
disc radius to be a free parameter, giving $r_{\rm out}=280$.
The overall continuum shape and luminosity is then fairly well
reproduced by   {\sc agnsed}   with $M=5.5\times 10^7~M_\odot$ and
$\dot{m}\simeq0.03$.
The best fit parameters are shown in table~\ref{tab:fit}, and the best fit model is overlaid on the 
decomvolved data points in Fig.~\ref{fig:fit}a.
  The estimated $\Gamma_{\rm warm}$ of 2.28
  is harder than the passive disc prediction of $2.5$ and may indicate 
  patchy corona as suggested by \cite{petrucci2017}. 
  We discuss this in more detail in Section 4.3.

  We also try to reproduce the data without any standard outer disc
  component, as in \cite{5548} and \cite{petrucci2017}. However, in our fit the seed photon
  energy is not a free parameter, but is set at the underlying disc
  temperature from reprocessing. The UV data are clearly in tension with
  this, as they show a stronger downturn than predicted by the warm
  Comptonisation models with seed photon energy set by the underlying
  disc area from reprocessing. A warm Comptonisation region covering the
  outer disc also changes some other aspects of the fit. 
The lower normalisation in the optical/UV (see
Fig.~\ref{fig:geometry}, geometry III) means that
  the same parameters of mass and mass accretion rate underpredicts  the
  optical data.
 Including the
  warm layer across all of the disc means that there should be an
  increase in  $(M\dot{M})^{2/3}$ to compensate for the decrease in normalisation from Comptonisation. 
  However, the observed bolometric luminosity $L_{\rm bol} =\eta
  \dot{M} c^2$ which is fairly well constrained by the data. Hence the only
  possibility to increase the optical emission is to increase the
  mass. This now pegs at the upper
  limit, giving a slight decrease in $\dot{m}\propto
  \dot{M}/M$. The outer radius is now not constrained from the data,
  with little change in goodness of fit from $ r_{\rm out}=10^3-10^5$
  so we set this back to the self gravity radius.  We show the best
  fit with these assumptions in Fig.~\ref{fig:fit}b,
  and tabulate the parameters in table~\ref{tab:fit}. 


In order to explain the optical data points which peak energy 
 at 8--10~eV with entire warm Comptonisation, blackhole mass needs to be as large as  $1\times 10^8~M_\odot$ with $\dot{m}\sim 0.01$ and $L_{\rm diss,hot}\sim 0.01L_{\rm Edd}$. 
This clearly exceeds the reasonable range of blackhole mass of NGC~5548. 
Hence we conclude  there is most likely a standard outer disc in NGC~5548.

\subsection{Mrk509: $\dot{m}\sim 0.1$}
\label{sec:mrk509}

Mrk 509 is the nearby Seyfert-1/quasar, and is one of the first objects in which
the soft X-ray excess was discovered \citep{singh1985}.  There was a
large multi-wavelength campaign on this
object \citep{kaastra2011-1,kaastra2011-2}, with simultaneous
optical-UV and X-ray monitoring from XMM-Newton's 
 OM and EPIC-pn together with
the HST/COS and archival 
observation by the Far Ultraviolet Spectroscopic Explorer (FUSE) \citep{509}.  As in
\cite{5548}, they fit the continuum SED without any outer disc
emission, just using a warm (0.2~keV) optically thick ($\tau\sim17$)
Comptonisation component with free seed photon temperature, together
with a hard ($\Gamma\sim1.9$) power-law X-ray continuum.

There is a clear iron line in the X-ray spectrum, together with a
Compton hump, so we include this in the model using {\sc gsmooth *
  pexmon} \citep{nandra2007}. 
The best fit model is shown in Fig.~\ref{fig:fit}c and detailed in table~\ref{tab:fit}.
The data are then well fit with our three component
{\sc agnsed}   continuum (i.e. including an outer disc) for a black hole
mass of $1\times10^8~M_\odot$ with $\dot{m}=0.1$.  
The transition radius from
the outer disc to warm Comptonised disc is around 
$r_{\rm warm}\simeq 40$, while the observed X-ray emission requires $r_{\rm hot}=21$. 

We also try a fit where the entire outer disc is covered by the warm Comptonisation (Fig.~\ref{fig:fit}d). 
This is statistically worse than the model with an outer standard disc, but unlike NGC~5548, the data match fairly well to the model around the UV peak. 
This model also gives $\Gamma_{\rm warm}\simeq 2.6$, consistent with a passive disc.
 \subsection{PG~1115+407: $\dot{m}\sim 0.4$}
\label{sec:pg1115}

PG~$1115+407$ is a NLS1 galaxy with mass from single epoch spectra of
$4.6\times 10^7~M_\odot$ using historical data with FWHM H$\beta$ of
1720 km/s \citep{porquet2004}, or $9.1\times 10^7~M_\odot$ with the
(narrow line subtracted) Sloan Digital Sky Survey (SDSS) FWHM H$\beta$ of 2310 km/s
\citep{jin2012a}. Hence we assumed a black hole mass range of 
(0.46--1.4)$\times 10^8~M_\odot$, including 0.2~dex uncertainty on the
SDSS limit (table~\ref{tab:system}).
We re-analyzed the same data set shown in \cite{jin2012a,jin2012b}, by concentrating on 
XMM/OM and XMM/EPIC-pn data alone since the SDSS data points 
were not simultaneous with the XMM observation
We fit the data with {\sc tbabs*redden*agnsed}, where {\sc tbabs} is used with 
\cite{anders1982} abundance and $E(B-V)$ is tied to 
$N_{\rm H}\cdot (1.7 \times10^{-22})$ as was done in \cite{jin2012a}. The observation is only 9.4ks, and the spectrum is 
steep.  This makes it difficult to constrain any reflection component so this is not included in the model.

\begin{figure*}
\includegraphics[angle=0,width=2\columnwidth]{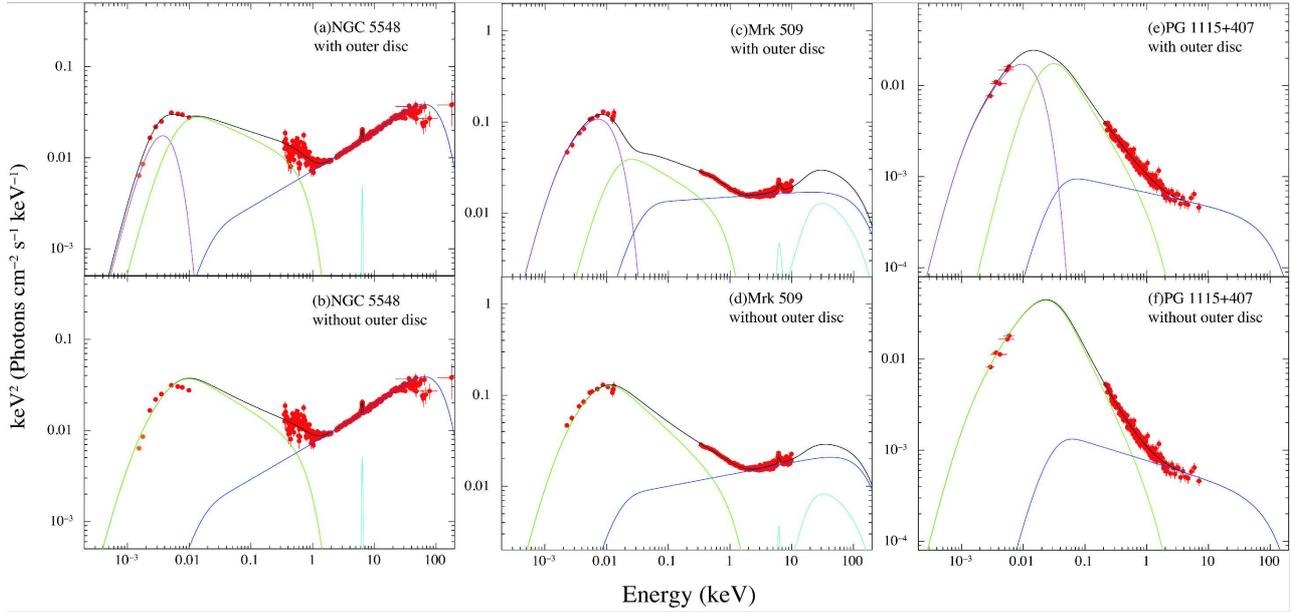}
    \caption{The best estimated models
    overlayd on the data set of NGC 5548 (\citealt[Fig. 10]{5548}), Mrk~509 (\citealt[Fig. 12]{509}), and PG~$1115+407$ \citep{jin2012a,jin2012b}.
    The outer disc emission, the warm compton component, and the hard compton component are shown in magenta, green, and blue, respectively. Panels (a), (b), (c), (d), (e) and (f) corresponds to NGC~5548 with and without outer disc, Mrk~509 with and without outer disc and PG~$1115+407$ with and without outer disc, respectively.}
\label{fig:fit}
\end{figure*}

As shown in table~\ref{tab:fit}, the model with an outer disc 
fits the data well, and $N_{\rm H}$ of $2.2\times 10^{20}~{\rm cm^{-2}}$ is consistent 
with galactic absorption of $(1.5$--$1.9)\times 10^{20}~{\rm cm^{-2}}$ \citep{kalberla2005,dickey1990}.
The unabsorbed SED derived from this best fit model is shown in Fig.~\ref{fig:fit}e.
Black hole mass is estimated as $M=1.0\times 10^8~M_\odot$ with
$\dot{m}=0.4$.
The size of the hot corona is $r_{\rm hot}=9.8$, which is much smaller  
than for the lower $\dot{m}$ AGN. 
For the warm Comptonising region, while 
$r_{\rm warm}$ is similar to that of Mrk~509, $\Gamma_{\rm warm}$ is much steeper at $\sim3.1$.
This most likely indicates that there is some intrinsic disc power dissipated 
underneath the warm corona rather than a completely passive disc. 
We compare this to models where the entire disc is dominated
by the warm Compton component. The fit results are shown in table~\ref{tab:fit} and Fig.~\ref{fig:fit}f.
The fit  is slightly worse in terms of $\chi_\nu^2$, and has slightly larger 
absorption at $N_{\rm H}=3.1\times 10^{20}~{\rm cm^{-2}}$.

\begin{table}
	\centering
	\caption{The best estimated parameters. $^\dagger$Electron temperature of the hot flow is fixed at 100~keV. $^\ast$Values in the parenthesis
	are internally calculated based on the other parameters.
	$^\ddagger$ The absolute values of $\chi^2 _\nu$ are not meaningful for deconvolved spectral fit. They are shown only for reference to compare the fit goodness  between with and without outer disc.Reflection components are modeled by a single gaussian and {\sc gsmooth*pexmon}   for  NGC~5548 and Mrk~509, respectively.
	$^{**}$ The values are limited by upper limit of parameters.}
	\label{tab:fit}
	\begin{tabular}{lccccc} 
		\hline
	&&
	NGC~5548&Mrk~509&PG$1115+407$\\ \hline\hline
\multicolumn{5}{c}{with outer disc}\\ \hline
$N_{\rm H}$ & ${\rm cm^{-2}}$ & --- & --- & $2.2\times 10^{20}$ \\
mass		&$M_\odot$&$5.5\times 10^7$&$1.0\times10^8$&$1.0\times10^8$\\
$\dot{m}$ &&$0.027$&$0.10$&$0.40$\\
$\Gamma_{\rm warm}$ &&2.28&2.36&3.06\\
$kT_{e,\rm warm}$ &keV&0.17&0.20&0.50\\
$\Gamma_{\rm hot}$& &1.60&1.96&2.14\\
$kT_{e,\rm hot}$ &keV&39&100$^\dagger$&100$^\dagger$\\
$L_{\rm diss,hot}$&$L_{\rm Edd}$&0.017&0.038&0.026\\
$L_{\rm hot}$&$L_{\rm Edd}$ &(0.018)&(0.042)&(0.040)\\
\multicolumn{5}{c}{..................................... size scales .....................................}\\
\multicolumn{2}{l}{$r_{\rm hot}$} &(43)$^\ast$&(21)$^\ast$&(9.8)$^\ast$\\
\multicolumn{2}{l}{$r_{\rm warm}$} &151&40&35\\
\multicolumn{2}{l}{$r_{\rm out}$} &282&(780)$^\ast$&(1416)$^\ast$\\
\multicolumn{5}{c}{.......................... characteristic temperatures ..........................}\\
$T(R_{\rm hot})$ &K&($2.9\times 10^4$&$5.4\times 10^4$&$1.0\times 10^5$)$^\ast$\\
$T(R_{\rm warm})$ &K&($1.3\times 10^4$&$3.7\times 10^4$&$5.6\times 10^4$)$^\ast$\\
$T(R_{\rm out})$ &K&($8.0\times 10^3$&$4.6\times 10^3$&$4.1\times 10^3$)$^\ast$\\
\hline
$\chi^2 _\nu$(dof)&&1.65(1097)$^\ddagger$&1.44(370)$^\ddagger$&1.64(184)\\ 
		\hline
		\hline
		\multicolumn{5}{c}{without outer disc}\\ \hline
$N_{\rm H}$ & ${\rm cm^{-2}}$ & --- & --- & $3.1\times 10^{20}$ \\
mass		&$M_\odot$&$6.0\times10^7$ $^{**}$&$3.0\times 10^8$ $^{**}$&$1.3\times 10^8$ \\
$\dot{m}$ &&$0.024$&$0.041$&$0.46$\\
$\Gamma_{\rm warm}$ &&2.39&2.59&3.35\\
$kT_{e,\rm warm}$ &keV&0.18&0.28&0.52\\
$\Gamma_{\rm hot}$& &1.61&1.87&2.21\\
$kT_{e,\rm hot}$ &keV&33&100$^\dagger$&100$^\dagger$\\
$L_{\rm diss,hot}$&$L_{\rm Edd}$ &0.016&0.013&0.021\\
$L_{\rm hot}$&$L_{\rm Edd}$ &(0.017)&(0.015)&(0.038)\\
\multicolumn{5}{c}{..................................... size scales .....................................}\\
\multicolumn{2}{l}{$r_{\rm hot}$} &(49)$^\ast$&(19)$^\ast$&(9.1)$^\ast$\\
\multicolumn{2}{l}{$r_{\rm warm,out}$} &$1000^{**}$&(407)$^\ast$&(1432)$^\ast$\\
\multicolumn{5}{c}{.......................... characteristic temperatures ..........................}\\
$T(R_{\rm hot})$ &K&($2.5\times 10^4$&$3.4\times 10^4$&$9.9\times 10^4$)$^\ast$\\
$T(R_{\rm warm,out})$ &K&$3.0\times10^3$&$4.4\times 10^3$&$4.0\times 10^3$)$^\ast$\\
\hline
$\chi^2 _\nu$(dof) &&2.11(1098)$^\ddagger$&2.01(371)$^\ddagger$&1.77(185)\\ \hline
	\end{tabular}
\end{table}

\section{Full AGN broadband spectral model}
\label{sec:summary}

In this section, we evaluate the results of 
fitting {\sc agnsed} to the 
observed SED of NGC~$5548$ ($\dot{m}\sim 0.03$), Mrk~509
($\dot{m}\sim 0.1$), and PG~$1115+407$ ($\dot{m}\sim 0.4$), and 
use these, together with other results in the literature, to 
build a full SED picture where the only free
parameters are $M$ and $\dot{m}$.

\subsection{Existence of an outer standard disc component}
\label{sec:outerdisk}

All the AGN in section~\ref{sec:application} are consistent with {\sc
  agnsed} of three components, where there is an outer disc, together
with warm Comptonising region and hot corona, powered by NT emissivity
for a low spin black hole. This is always a better fit than assuming
$r_{\rm warm}=r_{\rm out}$ i.e., a model where the warm Comptonisation
region extends over the entire outer disc,
although there are several uncertainties, e.g., on the inclination and 
absorption corrections.
Our model is different to that fit by \cite{5548} and \cite{petrucci2017}, where the optical/UV data are from the warm Comptonisation component alone.
This
difference is due to our assumption that the warm Comptonisation is
intrinsically linked to a NT disc. The data do fit just as well to an
unconstrained warm Comptonisation component as this has the same
optical/UV shape as a standard disc (see Fig.\ref{fig:comparison}
geometry III).
However, we have additional requirements on the luminosity and seed photon
temperature from our assumed NT emissivity.  The optically thick, warm
Comptonisation thus supresses the flux below that
predicted by the outer disc, so these models require a higher $M$ 
and/or higher absolute mass accretion rate, $\dot{M}$, but
the latter is fairly well constrained by the observed bolometric
luminosity from the broadband spectra. 

A larger $\dot{M}$
through the outer disc could fit the data if this is counteracted by
strong energy losses, e.g., if the system powers a UV line driven wind 
\citep{laor2014}. However, it seems
somewhat fine tuned that these wind losses (which vary with $M$ and
$\dot{M}$) would be always able 
to almost exactly compensate for the extra power 
predicted by
NT emissivity which assumes a constant $\dot{M}$ with radius. 
Similarly, high black hole spin would give higher
luminosity for a given $\dot{M}$ through the outer disc,
which can overpredict the total luminosity unless
this is mainly dissipated in the unobservable EUV bandpass. Again,
this seems fine tuned. The simplest solution is that we are seeing
evidence for an outer disc whose properties are like the standard
disc, and that the wind losses are small, and spin is low.

\subsection{Hot coronal emission}
\label{sec:hot}

As shown in table~\ref{tab:fit}, the observed power dissipated in the
hot inner flow is $L_{\rm diss,hot}=0.02$--$0.04L_{\rm Edd}$ in
our three AGN with different $\dot{m}$.  This is also seen in the
sample of 50 AGN in \cite{jin2012a}. These have different masses and
$\dot{m}$, but the X-ray luminosities are all the same within a
factor 2--3 when the SEDs are stacked into three groups of low,
medium and high $\dot{m}$ and referenced to the same black hole mass
(Fig. 8b in D12). 

\subsubsection{truncated disc and inner hot flow geometry}
\label{sec:truncated-disc}
\begin{figure}
\begin{center}
\includegraphics[angle=0,width=1\columnwidth]{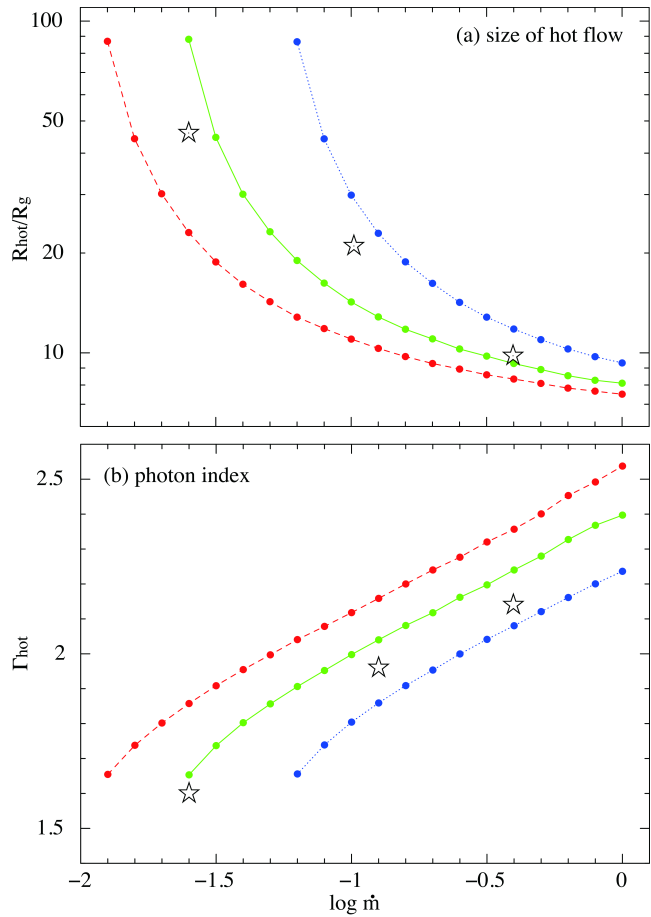}
\end{center}
    \caption{Radius of the hot inner flow, $r_{\rm hot}$ (a) and 
photon index of the hot Compton component, $\Gamma_{\rm hot}$ (b), are plotted against $\log \dot{m}$. 
The values are calculated assuming 
a fixed $L_{\rm diss, hot}$ of $0.01L_{\rm Edd}$ (dashed red line), $0.02L_{\rm Edd}$ (solid green line), and $0.05L_{\rm Edd}$ (dotted blue line). The spectral index is calculated with reprocess. 
  The observed values of $r_{\rm hot}$ and $\Gamma_{\rm hot}$ for 
 Mrk 509, NGC 5548 and PG~$1115+407$ are shown with open stars. 
}
    \label{fig:gamma}
\end{figure}

In our model, the approximately fixed value
for $L_{\rm diss,hot}$ measured from the data then determines $r_{\rm
  hot}$ from the NT emissivity.  For a total $\dot{m}\sim 0.03$, most
of the entire flow energy is needed to power the hard X-ray region,
thus $r_{\rm hot}$ is large. Conversely, for $\dot{m}\sim 0.4$, only a
very small fraction of the available power is needed to make the hard
X-ray flux, so $r_{\rm hot}$ is small. Figure~\ref{fig:gamma}a shows
how the value of $r_{\rm hot}$ decreases as a function of increasing
$\dot{m}$ in models where $L_{\rm diss,hot}$ is fixed at $0.01L_{\rm Edd}$
(dashed red), $0.02L_{\rm Edd}$ (solid green) and $0.05L_{\rm Edd}$
(dotted blue).  The open stars show the values
measured from fitting to the data scatter around the green line,
consistent with constant $L_{\rm diss,hot}=0.02L_{\rm Edd}$,
especially considering the uncertainties on mass determination.  The
predicted decrease in size scale of the X-ray source with increasing
$L_{\rm bol}/L_{\rm Edd}$ is compatible with the observations that the
X-ray variabilty timescale decreases with increasing $L_{\rm
  bol}/L_{\rm Edd}$ \citep{mchardy2006}.

We also calculate the self consistent spectral index, $\Gamma_{\rm
  hot}$, for the flow from eq.(14) in \cite{beloborodov1999} as

\begin{equation} 
\Gamma_{\rm hot}=\frac{7}{3}\left(\frac{L_{\rm diss, hot}}{L_{\rm  seed}}\right)^{-0.1}
\label{eq:gamma} 
\end{equation} 
With the geometry of
the truncated inner flow shown in Fig.~\ref{fig:geometry}a, $L_{\rm
  diss,hot}$ and $L_{\rm seed}$ are calculated via
eq.(\ref{eq:Lhotdiss}) and (\ref{eq:Lseed}), respectively.
Figure~\ref{fig:gamma}b shows the resulting $\Gamma_{\rm hot}$
assuming $L_{\rm diss, hot}$ at 0.01$L_{\rm Edd}$ (dashed red),
$0.02L_{\rm Edd}$ (solid green) and $0.05L_{\rm Edd}$
(dotted blue). The data (open stars) are in good agreement
with the truncated disc geometry for $L_{\rm diss, hot}=0.02L_{\rm
  Edd}$.
The observed spectral indices are then consistent with a truncated
disc geometry across the entire range of $\dot{m}$ considered here. 
This is surprising, especially at high $\dot{m}$, where the corona is more generally drawn as
either X-ray hot plasma over the inner disc, or as a lamppost (compact
source on the spin axis of the black hole). 
We explore untruncated disc geometries in section 4.2.2 and 4.2.3. 

\subsubsection{passive disc  and inner hot corona geometry}
Instead of the truncated disc and inner hot flow, there can be an inner disc corona extending down to $r_{\rm ISCO}$.
The inner disc corona can be characterised by the fraction, $f$, of
power dissipated in the 
corona with the remaining 
cooler disc material
in the midplane \citep{svensson1994}. 

We first assume a
maximal corona, with $f=1$ extending
over the inner disc from $r_{\rm hot}$ to $r_{\rm ISCO}$
i.e. a total power dissipated in the corona of $L_{\rm diss, hot}=0.02L_{\rm Edd}$
with a passive disc on the midplane.
The difference between
this and the truncated disc geometry (section \ref{sec:truncated-disc}) is that the inner disc on the
midplane will intercept half of the X-ray emission from the corona for
an isotropic source. The albedo, $a$, determines how much of the
illuminating flux can be reflected. The reflected fraction at low
energies depends strongly on the ionisation of the disc, but photons
above $\sim 50-100$~keV cannot be reflected elastically due to Compton
downscattering. This gives a maximum albedo for completely ionised
(most reflective) material and this value depends on spectral
shape. 
We evaluate this for a
Compton spectrum ({\sc nthcomp}) with $kT_{\rm
e}=100$~keV and different values of $\Gamma$ and
calculate the reflection albedo using {\sc ireflect} \citep{ireflect} with
$\xi=10^6~{\rm ergs~s^{-1}~cm}$. We choose this rather than the newer reflection models
such as {\sc relxill} as {\sc ireflect} calculates only the reflected
emission: the emission lines and recombination continuua will add to
the thermalised flux in making soft seed photons.
This gives $a_{\rm max}=0.55\sim 0.81$ for $\Gamma=1.5\sim 2.3$.

On the other hand, the seed photon power from reprocessing in the corona region results
in 
\[
\frac{L_{\rm diss,hot}}{L_{\rm seed}}=\frac{f}{1-\frac{1}{2}f-\frac{1}{2}fa}=\frac{2}{1-a}
\]
with maximal corona, $f=1$ (see eq.(3a) in  \citealt{haardt1993}).
This indicates 
\[
\Gamma_{\rm hot}=\frac{7}{3}\left(\frac{2}{1-a}\right)^{-0.1}
\]
via eq.(\ref{eq:gamma}). Thus, $a_{\rm max}$ and $\Gamma_{\rm hot, min}$ is self-consistently determined as $a_{\rm max}\sim0.7$ and 
$\Gamma_{\rm hot, min}\sim 1.9$ for an inner disc corona geometry.
In Fig.~\ref{fig:gamma_slab}, which shows the truncated
disc results as a baseline model (green solid line) together with
the data (open stars) from Fig.~\ref{fig:gamma},  
we plot $\Gamma_{\rm hot}$ as a horizontal dotted black line.
This horizontal dotted black line shows this minimum photon index.
The observed $\Gamma_{\rm hot}$ for Mrk~509 sits on this lower limit, 
so can be explained also by this geometry. 
PG~$1115+407$ is somewhat steeper at $\Gamma_{\rm hot}=2.2$, which can also be explained by this geometry 
for $a=0.3 (<a_{\rm max})$.
However, NGC~5548 and other AGN with $\Gamma_{\rm hot}<1.9$ require a more photon starved geometry.

\subsubsection{entire slab hot corona}
\begin{figure}
\begin{center}
\includegraphics[angle=0,width=1\columnwidth]{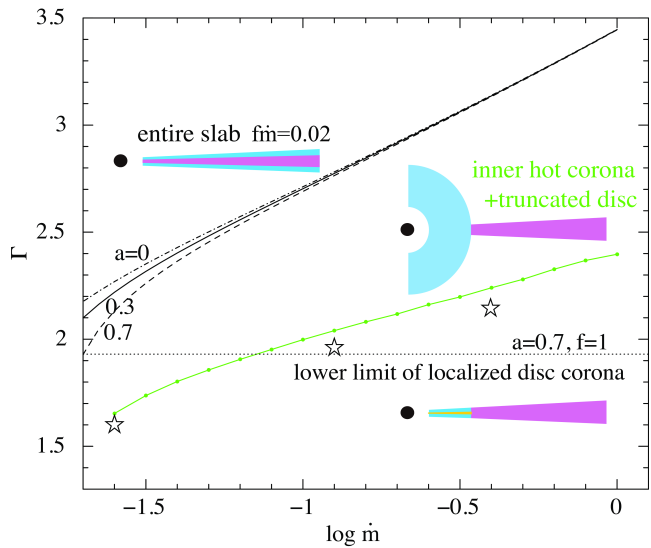}
\end{center}
    \caption{Same as Fig.~\ref{fig:gamma}b, but expected values of $\Gamma_{\rm hot}$ for slab geometry 
    with $L_{\rm diss,hot}=0.02L_{\rm Edd}$ (i.e., $f\dot{m}=0.02$ with the fraction of power dissipated in the corona, $f$) are shown in black lines, together with the observed values of $\Gamma_{\rm hot}$ for 
 Mrk 509, NGC 5548 and PG~$1115+407$ (open stars) and $\Gamma_{\rm hot}$ for truncated disc geometry with $L_{\rm diss,hot}=0.02L_{\rm Edd}$ (solid green). Albedo is assumed to be $a=0$ (dash-dot), 0.3 (solid) and 0.7(dash).
Horizontal dotted straight lines represent the lower limit of $\Gamma$ under an assumption of  localized slab corona with maximal albedo, $a=0.7$ and by assuming 'passive disc' underneath the corona, i.e., $f=1$.
}
    \label{fig:gamma_slab}
\end{figure}

The value of $\Gamma_{\rm hot}$ becomes larger if there is  intrinsic disc power (i.e., $f<1$), which adds to the seed photons
in the local slab geometry. This requires larger 
$r_{\rm hot}$  to keep the same 
observed $L_{\rm diss, hot}/L_{\rm Edd}$.
The most extreme case is where the corona extends over the entire optically thick disc
so that $f$ is constant with radius. 
For such a entire slab geometry, 
the seed photons are given by  eq.(3a) in  \cite{haardt1993} as
\[
\frac{L_{\rm seed}}{L_{\rm Edd}}= \left(1-\frac{1}{2}f-\frac{1}{2}fa\right)\dot{m}
\]
In this geometry the  hard X-ray dissipation of  $L_{\rm diss,hot}/L_{\rm Edd}=f\dot{m}=0.02$ 
implies $f=0.02/\dot{m}$. We use this in the equation above to
calculate $L_{\rm diss, hot}/L_{\rm seed}$ as 
\[
\frac{L_{\rm diss,hot}}{L_{\rm seed}}=\frac{f\dot{m}}{ (1-\frac{1}{2}f-\frac{1}{2}fa)\dot{m}}=\frac{2}{100\dot{m}-1-a}
\]
and hence derive the spectral index via eq.(\ref{eq:gamma}).
We plot these in Fig.~\ref{fig:gamma_slab} for three values of the
albedo namely $a=0$ (dotted), $0.3$ (solid) and $0.7$
(dashed).  These always give much steeper $\Gamma_{\rm hot}$ than observed. 
Figure~\ref{fig:gamma_slab} shows that steep spectra of $\Gamma_{\rm hot}\simeq 3.1$--3.4 
are predicted for $\log \dot{m}=-0.6\sim0$ while the observed photon indices are usually harder as 
$1.7\sim 2.4$ (e.g., from 55 samples in \cite{jin2012a}.
For lowest $\dot{m}$, $\Gamma_{\rm hot}$ is still as large as 1.9.
Thus the entire slab for the hot Comptonisation component is quite unlikely for both high and low $\dot{m}$.
This problem is discussed in detail also e.g., 
by \cite{stern1995}, \cite{malzac2005} and \cite{poutanen2017} for the case of BHBs in the low/hard state. 
The problem is even more marked for AGN as the disc seed
photons are at lower energies, making Compton cooling more efficient
and leading to steeper predicted photon indices \citep{haardt1993}.

\subsubsection{summary of constraints on the geometry of the hot corona}
To summerize, the data at $\dot{m}\le 0.1$ have $\Gamma_{\rm hot} <1.9$ which
is incompatible with a disc-corona geometry even in the limit where all the power is dissipated in the corona. Spectra at higher $\dot{m}\ge 0.1$ have spectra which can be produced in an inner disc-corona geometry, but these require some fine tuning in order to produce the spectral indices observed.
By contrast, the truncated disc/hot inner flow geometry can predict
the behaviour of the spectral index over the entire range of $\dot{m}$,
making it likely that this geometry is continued across all $L_{\rm bol}/L_{\rm Edd}$.

\subsection{Warm comptonisation region}

The warm Comptonisation region extends from $r_{\rm hot}$ to an outer
radius $r_{\rm warm}\lesssim r_{\rm out}$.  Ideas which associate this with the
changing vertical structure of a disc due to the importance of atomic
opacities predict that the warm Comptonisation region should onset at
an approximately fixed temperature.  One attractive idea is that each
annulus of the disc which is at the same temperature as an O star
would be modified by a UV line driven wind \citep{laor2014}. The
maximum disc temperature considering these wind losses is (2--8)$\times
10^4$~K in their models, which is similar to the range seen here for
the onset of the warm Comptonisation of (1--6)$\times 10^4$~K.  Thus
it is possible that the onset of warm Comptonisation does link to the
changing disc structure induced by UV opacities, though strong wind
losses such as those predicted by \cite{laor2014} are ruled out by the
observed efficiencies. 

This overprediction of wind losses is probably
linked to the assumption in \cite{laor2014} that there is only a disc,
with no hard X-ray emission which will strongly suppress UV line
driving by overionisation \citep{proga2000}. It seems plausible that
the UV bright disc launches a UV driven disc wind, but that this
becomes ionised as it rises up and is exposed to the X-ray source. The
failed wind falls back down, impacting the disc, leading to shock
heating of the photosphere, making the warm Comptonisation region. 
There are no  calculations of this at the current time, but we expect 
that the extent of the failed wind will depend on the level of X-ray
ionisation. This is clearly larger for lower $\dot{m}$. Guided
by the fits to the individual objects above, we tie the 
size scale, $r_{\rm warm} =2r_{\rm hot}$.

While the data do not favour all the outer disc being
covered by the warm corona, they are (mostly) consistent with the idea
that the disc underneath the warm Comptonising material is passive
i.e., the optically thick material underneath the corona only
reprocesses the luminosity dissipated further up in the disc \citep{petrucci2017}.
%
There is a tred seen also in \citep{petrucci2017},
$\Gamma_{\rm warm}$ is somewhat steeper for high $\dot{m}$, 
(e.g., PG1115+407 indicates some intrinsic power in the disc),  
and somewhat flatter at low $\dot{m}$ (e.g., NGC~5548).
Steeper spectra can easily be produced with some intrinsic emission 
on the disc midplane.
However, the lower spectral indices at low
$\dot{m}$ are more difficult to explain.

 \cite{petrucci2017} suggest
that $\Gamma_{\rm warm}\le 2.5$ 
is from partial covering of the corona over the passive
midplane, so that some of the reprocessed photons do not re-intercept
the warm Comptonisation region to cool it.  
While this does indeed allow reprocessed photons to escape, this also means that these
seed photons from the disc are seen directly which is not consistent with their assumption that
the optical/UV is dominated by warm Comptonisation alone, with no
thermal emission from a disc (see also the discussion of this in 
\citealt{petrucci2017}).  Partial covering also seems physically unlikely as the optically thick, warm material
has thermal pressure so should expand outwards so this additionally requires 
magnetic confinement. 
Instead, we suggest that the harder spectral indices could instead be produced by
irradiation heating being more important at low $\dot{m}$ (see
also \citealt{lawrence2012}). By definition, irradiation heats the
photosphere at $\tau=1$ rather than the deeper regions at $\tau=10-20$
where the majority of the warm Comptonisation must be produced. We
suggest that a more accurate treatment of an irradiated warm
Comptonisation region above a passive disc could produce the harder
indices observed at low $\dot{m}$.

\begin{table}
	\centering
	\caption{Parameters used in section \ref{sec:uv-x} and \ref{sec:reprocess}. $^\dagger$Geometry of hot inner flow + warm comptonising skin + outer disc. $^\ddagger$Geometry of hot inner flow + warm comptonising skin (i.e., without outer disk). }
	\label{tab:parameter}
	\begin{tabular}{llcccc} 
		\hline\hline
system parameter&$r_{\rm out}$& $r_{\rm sg}$\\
&inclination angle &$45^\circ$\\
hot inner flow&$T_{e, \rm hot}$&100~keV\\
&$\Gamma_{\rm hot}$&calculated via eq.(\ref{eq:gamma})\\
&$L_{\rm diss, hot}$&$0.02L_{\rm Edd}$\\
warm compton&$T_{\rm warm}$&0.2~keV\\
&$\Gamma_{e, \rm warm}$&2.5\\
&$r_{\rm warm}$&$2r_{\rm hot}~^\dagger$, $r_{\rm out}~^\ddagger$\\
reprocess&&included\\
\hline
	\end{tabular}
\end{table}

\section{Predictions of the full SED model for the observational results}
\subsection{UV/X relation}
\label{sec:uv-x}

We use the individual AGN fits to define the full SED model in order
to compare to a large sample of objects spanning a wide range in
$\dot{m}-M$. \cite{lusso2017} show that there is a well
defined relationship between the UV and X-ray luminosity of AGN, and
claim this has low enough scatter to be used as a tracer of the cosmological
expansion. Finding the underlying physics is then extremely important,
as it can only be used with confidence when it is robustly understood.

Guided by the individual AGN fits above, we fix 
$L_{\rm diss, hot}=0.02L_{\rm Edd}$, which defines $r_{\rm hot}$. We assume that the hot
inner flow is quasi-spherical, and that the optically thick flow
truncates at $r_{\rm hot}$ so that the spectral index, $\Gamma_{\rm hot}$, can
be calculated from the ratio of dissipation in the hot region to
intercepted seed photons. The best fit models above strongly support a
geometry where there are three components, but again we compare the
data to a model where the entire outer disc is covered by the warm
Comptonising material. 
The inclination angle is expected to be between $0^\circ$ and
$60^\circ$ for type 1 AGN so we fix this at $45^\circ$.
Full model parameters are shown in table~\ref{tab:parameter}. 
The model with $r_{\rm warm}=2r_{\rm hot}$ defines our simplified model {\sc qsosed}. 

\begin{figure*}
\includegraphics[angle=0,height=0.7\columnwidth]{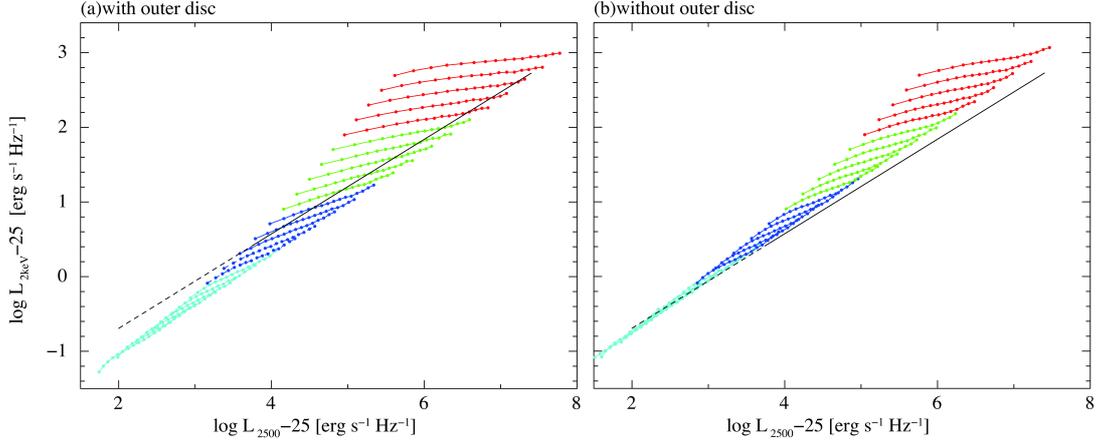}
    \caption{Monochromatic luminosities log $L_{\rm X}$ against log $L_{\rm UV}$. 
   for black hole of $M=(0.1$--$1)\times10^7M_\odot$ (cyan), $(0.16-1)\times 10^8~M_\odot$ (blue), 
   (0.16--$1)\times 10^9~M_\odot$ (green), 
   (0.16--$1)\times 10^{10}~M_\odot$ (red). 
   From left to right $\dot{m}$ changes from 0.03 to 1.
    The observed UV/X relation in the range of $\log L_{\rm
      2500}-25=$3.8--7.4\citep[Fig. 3]{lusso2017} is shown with a
    solid line. A dashed line is an extrapolation of the solid line.
(a) Our three component flow, with an outer disc,
warm Comptonisation region and hot inner flow. (b) A
model where there is no standard outer disc. 
}
    \label{fig:uv-x}
\end{figure*}

We calculate a grid of models spanning $\dot{m}=0.03$--1 and
$M=10^6$--$10^{10}M_\odot$. 
This is consistent with the range of $\dot{m}$ and $M$ in \cite{lusso2017} 
(the majority of their sample have $\dot{m}=0.03$--1 and (1--10)$\times 10^8M_\odot$; \citealt{lusso2012}). 
Figure~\ref{fig:uv-x}a shows the resulting relation between the
monochromatic rest frame luminosity at 2500\AA~(i.e., 5 eV) and 2~keV, with lines
connecting varying $\dot{m}$ for constant $M$ from our model
with an outer standard disc. 
Figure~\ref{fig:uv-x}b shows the slightly different predictions from a
model where the warm Comptonisation region covers the entire outer
disc. Both these give a fairly well defined correlation 
between the UV and X-rays, though with some scatter. 
We compare these to
the best fit UV/X relation derived from 545 SDSS quasars by
\cite{lusso2017} of
\begin{equation}
\log L_{\rm 2keV}-25=0.633 \cdot (\log L_{\rm 2500 \AA}-25)-1.959
\end{equation}
(solid black line) on Fig.~\ref{fig:uv-x}a and b. It is clear that 
the observed correlation is a slightly different slope than predicted
by our models over this entire range. Our models with an outer disc match quite well to the observed relation at
high masses (Fig.~\ref{fig:uv-x}a), while models with just a warm Comptonisation region are
offset by their smaller predicted UV luminosity and match better at
low masses (Fig.~\ref{fig:uv-x}b). 

Our predicted correlation is easy to understand, as it 
arises due to our assumptions that the X-ray flux is fixed at
$L_{\rm X}=0.02L_{\rm Edd}\propto M$ while the UV is from a disc/comptonised
disc so $L_{\rm UV}\propto (M\dot{M})^{2/3}\propto (M^2\dot{m})^{2/3}$.
Thus this predicts $\log L_{\rm X}=\frac{3}{4}\log L_{\rm UV} -
\frac{1}{2}\log\dot{m}+b$ where $b$ is a constant. There is clearly
scatter in the models from $\dot{m}$, but the range should be 
constrained between $0.02-1$. 
The lower limit is where the 
entire accretion flow
is expected to make a transition to an ADAF. \cite{narayan1995}, so that there is no bright UV 
emitting disc left, so that there is no source of ionising flux to excite a BLR.
The upper limit comes from the expectation that 
super-Eddington objects are rare.
Thus there is only a limited range of 1.5~dex in $\dot{m}$, and
this reduces to 0.75~dex with the square root factor. However, the
data from \cite{lusso2017} have a scatter of only 0.2~dex.

\begin{figure*}
\includegraphics[angle=0,height=0.7\columnwidth]{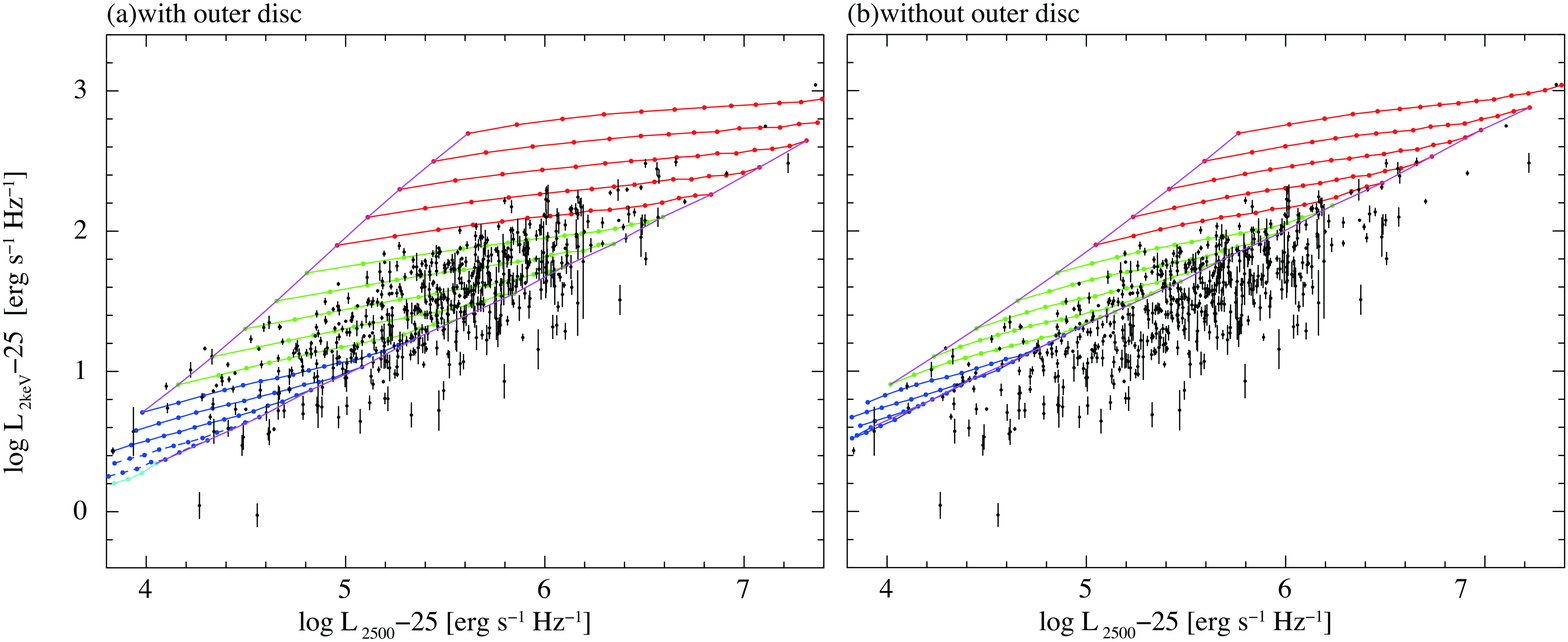}
    \caption{Enlargements of Fig.~\ref{fig:uv-x} are overlayed on the observed data points in Fig. 3 of \citet{lusso2017} shown with open grey circles.}
    \label{fig:uv-x_lusso}
\end{figure*}

We study the predicted behaviour in more detail in
Fig.~\ref{fig:uv-x_lusso}, using the individual data points from
\cite{lusso2017} (B. Lusso, private communication). These are selected
from the SDSS quasar sample, so have absolute i-band magnitude
brighter than $-22$. This already limits the black hole mass to
$>10^{7.5}M_\odot$ for objects with a standard disc below
Eddington, and the masses reported in \citep[their Fig.6]{lusso2012} are 
clustered around $(1-10)\times 10^8~M_\odot$. 
Thus their data only include high black hole masses, so our models
with an outer standard disc match fairly well in normalisation and slope
to that observed (Fig.~\ref{fig:uv-x_lusso}a), whereas models with
complete coverage of the outer disc with a warm Comptonisation
underpredict the UV luminosity (Fig.~\ref{fig:uv-x_lusso}b).

Nonetheless, there is still a mismatch between the range predicted by
our models and the observed data, even including an outer
standard disc. The data extend to slightly higher UV luminosity than
expected for Eddington limited systems. Some of this can be inclination, as
more face on systems will have stronger UV flux, but this makes only a difference
of $0.15$ dex between our assumed mean inclination of $45^\circ$ and $0^\circ$. 
While this may explain the AGN not covered by the grid in the model with an outer standard
disc, it is not enough to explain the larger number missed by the grid if the warm 
corona covers the entire outer disc. Instead, these require a substantial 
population of super-Eddington AGN. Super-Eddington AGN are seen in the local Universe
\citep{jin2015,done2016,jin2017}, though they are rare, but their high
UV luminosity enhances their probability of selection.
We will extend the model to super-Eddington flows in a later work (Kubota \& Done, in preparation), but note here that there are multiple uncertainties with the structure of these flows which makes robust prediction difficult. 

There is the opposite problem for the highest mass black holes at the lowest
$\dot{m}$, where the grid extends into a region where there are no data points. 
We suggest that this could be due to selection
effects.  High mass black holes are rarer, so require sampling a
larger space volume in order to have a realistic probability of
finding some. This means that they are generally seen at larger
distances, so are only selected in flux limited samples if they have
high luminosity, weighting the selection of high mass black holes
towards higher $\dot{m}$ (including super-Eddington rates).

Thus the observed UV/X-ray relation is predicted by our model, where the
X-ray lumininosity is fixed at the $0.02L_{\rm Edd}$ and the
UV is from an outer standard disc, with selection effects (mostly the
limited range of black hole mass) suppressing
some of the predicted scatter. This is a very different explanation to
that of \cite{lusso2017}. Their 'toy' model uses the same standard
disc equations to estimate the UV flux, but they set the X-ray flux
using the gravitational power emitted from the outer disc down to the
radius at which the disc becomes radiation pressure dominated. As they
note in their paper, producing the X-rays at large radii in the disc
rather than close to the black hole is in conflict with microlensing
size scales, as well as with the rapid X-ray variability. We suggest
that their model works because it effectively hardwires $L_{\rm X}$ to a
constant value. The radius at which radiation pressure dominates in
the disc, $R_{\rm rad}$, increases as $\dot{m}$ increases, so leaving a
smaller and smaller fraction of power dissipated in the outer disc,
and hence reproducing the observed decrease of $L_{\rm X}/L_{\rm bol}$ with
$\dot{m}$ \citep{vasudevan2007}. 

Formally, this gives $R_{\rm rad} \propto (\alpha
M)^{2/21} \dot{m}^{16/21}(1-f)^{6/7} R_g$, where $f$ is the fraction
of the accretion power which is dissipated in the hard X-ray corona
\citep{svensson1994}. Then the X-ray luminosity down through the
corona from $R_{\rm out}-R_{\rm rad}$ is 
\[L_{\rm X}\propto fGM\dot{M}/R_{\rm rad}\propto
\alpha^{-2/21}M^{19/21}\dot{m}^{5/21}f(1-f)^{-6/7}
\] (see their eq.(14) in detail).  Hence $L_{\rm X}$ is roughly
proportional to $M$ and has only a weak dependence on
$\dot{m}$, so their toy model is almost identical with our
assumption of constant $L_{\rm diss,hot}=0.02 L_{\rm Edd}$.
 
Our model hardwires the same absolute value of $L_{\rm X}$, but in a much
more plausible geometry where the X-ray source is produced close to
the black hole, and with more physical motivation.

\subsection{Optical variability}
\label{sec:reprocess}

The X-rays vary
rapidly in a stochastic manner about a mean, so their reprocessed
emission should also carry the imprint of this rapid variability. 
The assumption in our SED is that the hard X-rays carry a fixed
luminosity but arise from a smaller region as $\dot{m}$ increases. 
Hence the reprocessed luminosity 
depends on $L_{\rm diss, hot} \times r_{\rm hot}$ (see Section 2.3) 
which decreases as  $\dot{m}$ increases. 
This reprocessed flux is seen against 
the remaining, constant component from the disc and/or warm Comptonisation region 
which increases with $\dot{m}$. Thus the variable
reprocessed emission forms a smaller fraction of the optical/UV
emission at higher $\dot{m}$, which qualitatively matches to what is
observed \citep{macleod2010,ai2013,simm2016,kozlowski2016}.  

\begin{figure}
\begin{center}
	\includegraphics[angle=0,width=0.95\columnwidth]{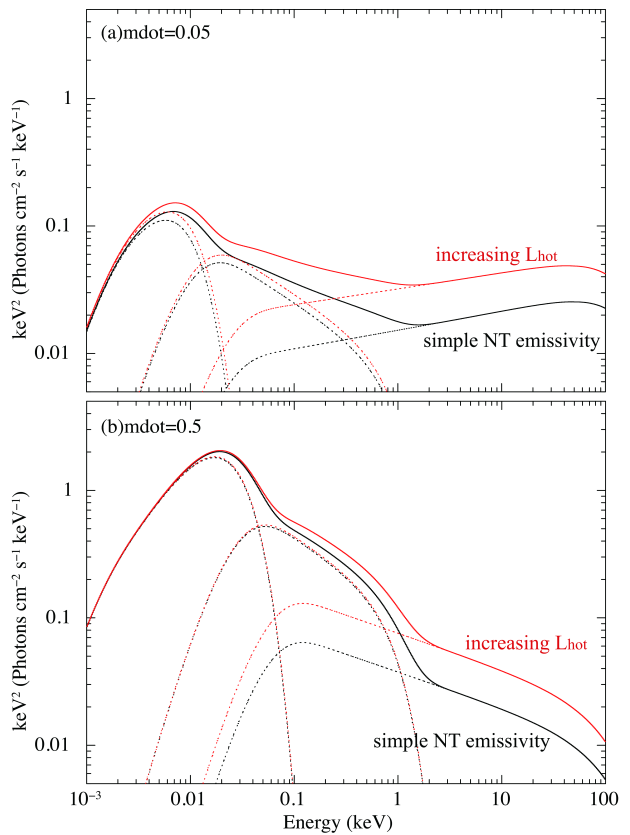}	
\end{center}
\caption{Effects of hard X-ray reprocess
 for a black hole of $M=10^8~M_\odot$ with $\dot{m}=0.05$ (a) and 0.5 (b).
SEDs in which the hard X-ray luminosity is increased by a factor of 2.0 (red) are compared with those with Novikov-Thorne emissivity with hard X-ray reprocess (black).
}
\label{fig:reprocess_time}
\end{figure}

Our model explicitly includes the reprocessed emission from X-ray
illumination of the outer disc and warm Comptonisation region, so here
we calculate the contribution that the varying X-ray emission can make
to the optical variability.  We illustrate this with our three
component SED model, {\sc agnsed}, (with all parameters as above, tabulated in 
table~\ref{tab:parameter} with $r_{\rm warm}=2r_{\rm hot}$) for AGN of $10^8M_\odot$ with
$\dot{m}=0.05$ and 0.5 in Fig.~\ref{fig:reprocess_time}a and b,
respectively.  The black lines show spectra based on the simple
NT emissivity including reprocessing, while red lines are the result of
stochastic variability (so no impact on $\dot{m}, r_{\rm hot}$ etc) 
increasing $L_{\rm hot}$ by a factor of 2.  The
emission from both the outer disc and warm Comptonising region
increases with the increase in $L_{\rm hot}$, and it is clear that
reprocessing makes a much larger fraction of the optical emission at
low accretion rates than at high ones.

\begin{figure}
\begin{center}
\includegraphics[width=1.1\columnwidth ]{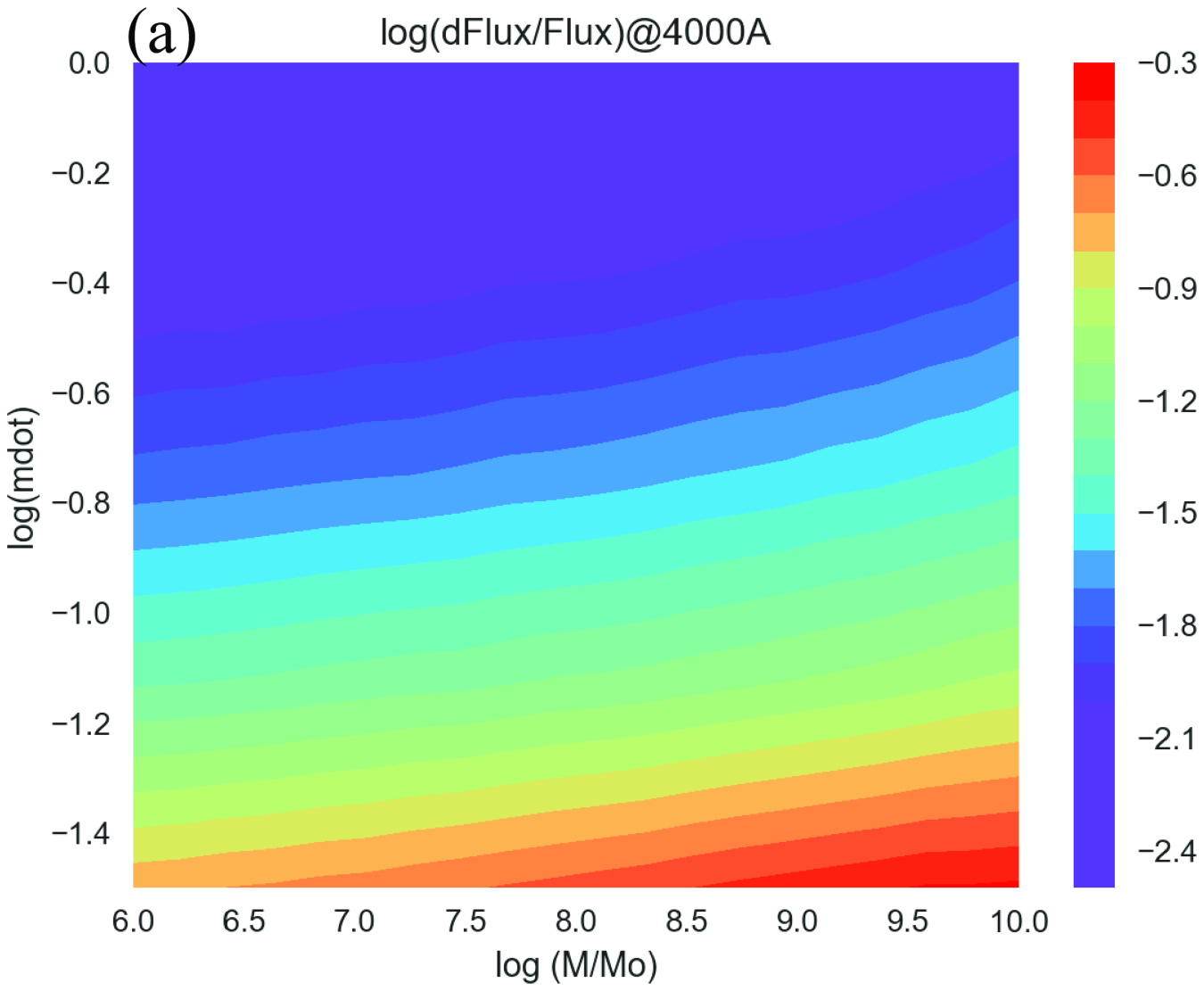}
\includegraphics[width=1.1\columnwidth ]{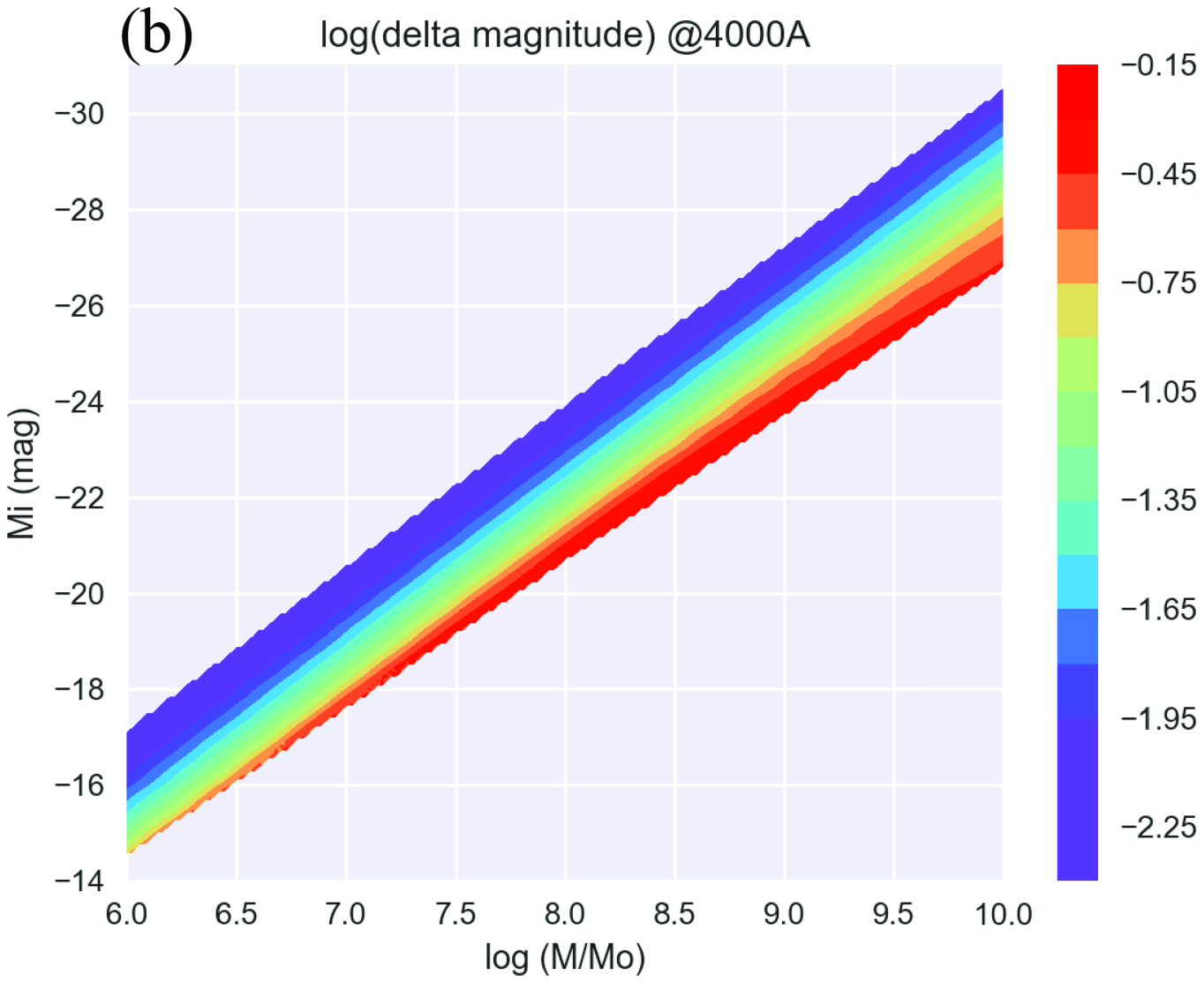}
\end{center}
    \caption{(a)The time variability $\log \Delta f_{4000}/f_{4000}$ are shown as color grid of 
    black hole mass $\log M/M_\odot$ and  $\log \dot{m}$. Hot X-ray emission is increased by a factor of 2.0. 
    (b) Same as the top panel but $\log \Delta flux/flux$ and $\log \dot{m}$ are converted into 
    $\log \Delta mag$ at 4000\AA~ and $i$-band absolute magnitude $M_i$. SEDs are based on the parameters of model (1) in table~\ref{tab:parameter}.}
   \label{fig:variability}
\end{figure}

We quantify the fractional change in optical flux at 4000\AA~
(3.1~eV), $\Delta f_{4000}/f_{4000}$, to a factor 2 increase in
$L_{\rm diss,hot}$ across the entire range of AGN masses ($M=10^6\sim
  10^{10}M_\odot$) and mass accretion rates ($\dot{m}=0.03$--1). 
Figure~\ref{fig:variability}a shows
this fractional variability as the colour coding across the grid of
$\log M/M_\odot$ and $\log \dot{m}$.  It is obvious that lower
$\dot{m}$ gives larger variability, though there is also a much
smaller effect where the variability increases with larger mass. This
occurs when reprocessing starts to affect the variability at the peak
of the warm Comptonisation region, where the temperature shift
increases the amplitude of variability (see Fig.~\ref{fig:reprocess}).

We mach this to the observed amplitude of variability seen in a large
sample of quasars in SDSS stripe 82.  \cite{macleod2010} characterised
the variability at rest frame 4000\AA~ with a damped random walk.  The
mean asymptotic amplitude of variability in optical magnitudes
is characterised by the structure function extrapolated
to infinite time, $SF(\infty)$. This is plotted as a function of
$i$-band magnitude, $M_i$, and black hole mass in their Fig.~14. In
order to compare our models directly to their results, we convert our
mass and $\dot{m}$ to $M_i$, and  convert our fractional
variability at 4000\AA~ to a magnitude difference ( $2.5 \Delta \log
f_{4000} = \Delta m $). Figure~\ref{fig:variability}b shows 
this magnitude
difference as a function of black hole mass and $M_i$ across the whole
range of models calculated in Fig.~\ref{fig:variability}a. 

\begin{figure}
\begin{center}
\includegraphics[width=1.1\columnwidth ]{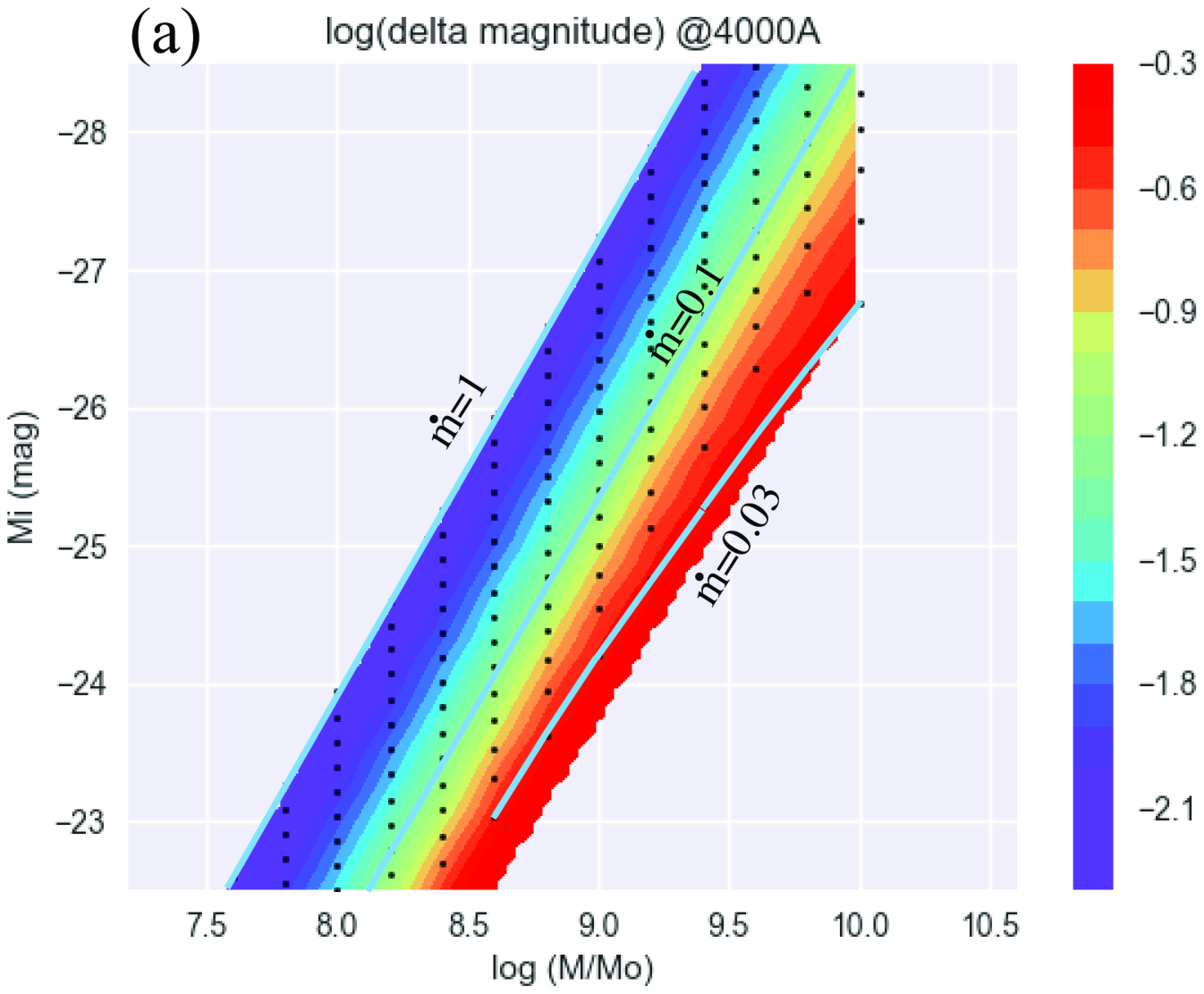}
\includegraphics[width=1.1\columnwidth ]{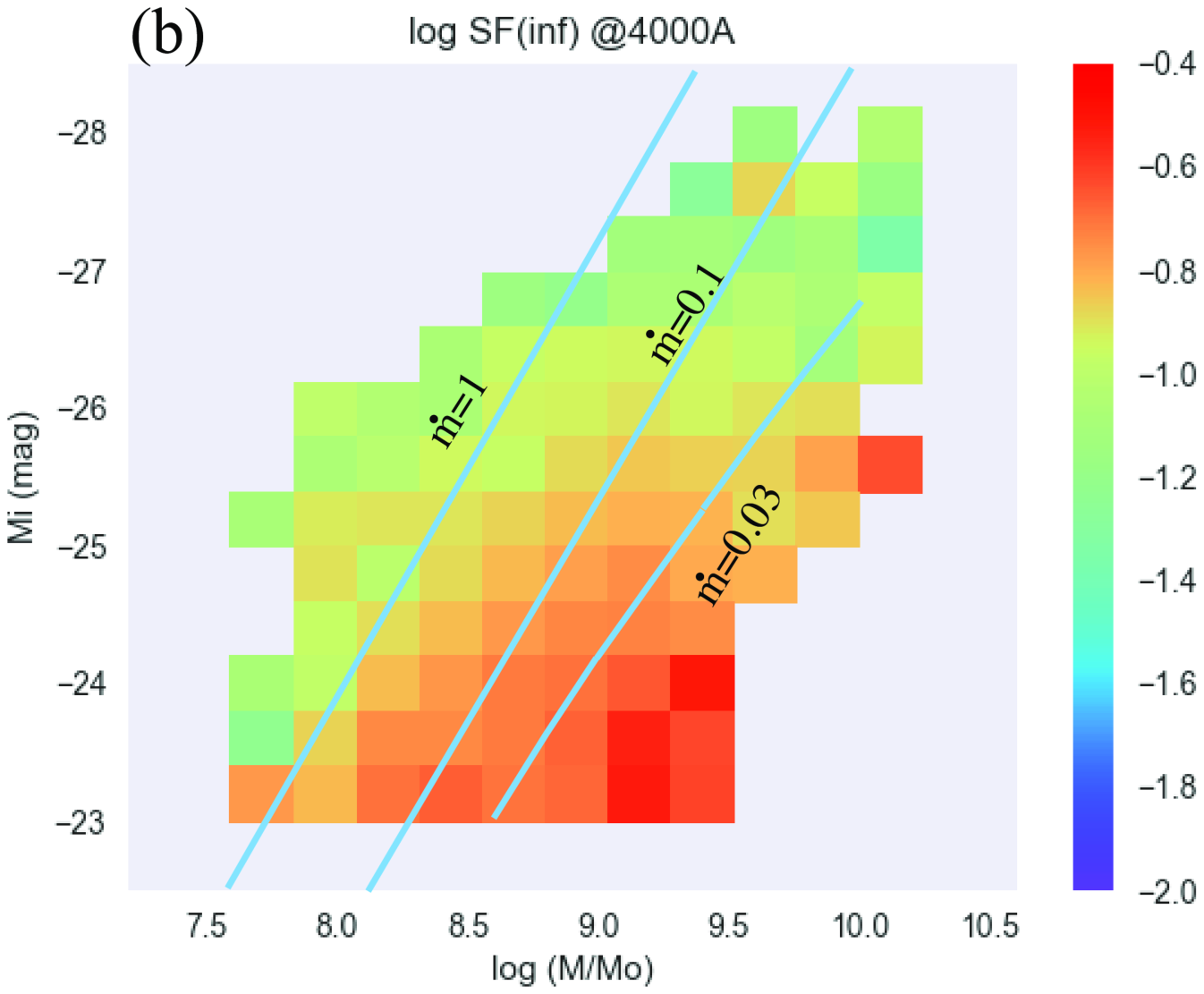}
\end{center}
    \caption{ (a) Enlargement of Fig.~\ref{fig:variability}b. Black dots represent the places which we calculate.  
    (b) $\log SF(\infty)$ at 4000\AA~ are plotted 
    in space of $M_i$ and $\log M/M_\odot$(Fig.14 in \citealt{macleod2010}). These data points are given by
 C.Macleod in private. Colour grid is the same between these two panels.
 Contours of  constant $\Delta mag$ in the top panel are overlaid on the bottom panel. Solid cyan lines show 
 constant $\dot{m}$ of 0.03, 0.1 and 1.}
   \label{fig:macleod}
\end{figure}

Figure~\ref{fig:macleod}a shows a zoom of Fig.~\ref{fig:variability}b,
limiting it to the same range in mass and $M_i$ as used by
\cite{macleod2010}. This can then be compared directly against the
data in Fig.\ref{fig:macleod}b (C. Macleod in private
communication). 
The range in $M_i$ for each black hole mass spanned by our models at a
given $\dot{m}$ is shown by the cyan lines for 
$\dot{m}$ of 0.03, 0.1 and 1. 
 This makes it plan that the
most variable AGN are those with implied  $\dot{m}<0.03$. These 
plausibly connect to the 'changing look' quasars if
these are triggered by a state change from an outer disc to an ADAF
flow (e.g.,\citealt{noda2018}). 

Figure~\ref{fig:macleod}a also stresses the need to use the non-linear
conversion between optical flux and bolometric luminosity which is
inherent in the standard disc equations. Hence our data 
with $\dot{m}=0.03-1$ do not span
the entire range of $M_i$ which are associated with this range in
$\dot{m}$ in Fig.15 of \cite{macleod2010}.
There are again some AGN with 
$\dot{m}>1$. These are rare (Fig.~12 in \citealt{macleod2010}), but
they considerably extend the range in $M_i$ for the lowest mass AGN
included here. 

The colour grid is the same between Fig.~\ref{fig:macleod}a and b, and
the models with a factor 2 variability in X-rays give the observed
amount of optical variability at $\dot{m}=0.03$.  Our models predict a 
weak trend for higher variability at higher black hole masses at fixed $\dot{m}$, which
is opposite to the observed weak trend for higher variability at lower
$M$ but this may not be a serious discrepancy as the timescales for
the higher mass black holes are longer, which lead to some
variability being missed. A much bigger discrepancy is that the models
predict almost no variability at $\dot{m}=1$, unlike the data which
still show variability at the 10\% level.  This clearly shows that
some other component is required to make at least part of the optical
variability, though the most rapid variability (e.g. from Kepler light
curves: \citealt{aranzana2018}) must arise from X-ray reprocessing. 
An additional source of longer term optical variability also 
matches with results from more intensive monitoring
campaigns which stress the lack of correlation between the X-ray and
optical lightcurves (e.g., \citealt{arevalo2009}). 
\cite{gardner2017} suggest that variability of the soft X-ray excess may play a role, but
in our models here this still makes little impact on the optical spectrum at
$\dot{m}=1$. Instead, there should be intrinsic variability in the
disc spectrum assuming our disc dominated SED are the correct
description of AGN close to the Eddington limit. Standard disc models
do indeed predict that AGN discs should be highly unstable due to
their dominant radiation pressure, but the non-linear outcome of what
should be limit cycle variability is not yet known
(\citealt{hameury2009} use a heating prescription which goes with only
gas pressure to avoid complete disc disruption). 

\section{Summary and Conclusions}

We construct a new spectral model {\sc agnsed}   which includes an outer
standard disc, a middle region where the disc is covered by optically
thick, warm Comptonising electrons, and an inner region of hot plasma
which emits the power law X-ray component. We assume that these
regions are separated in radius, and that their emission is determined
by the overall NT emissivity. This sets the size scale of
the hot X-ray plasma and we include reprocessing of the X-rays from
this source which illuminate the outer and warm Comptonising disc.

We fit this model to multiwavelength SEDs of three well observed AGN
with very different Eddington ratios, NGC~5548 ($\dot{m}\sim 0.03$),
Mrk~509 ($\dot{m}\sim 0.1$), and PG~$1115+407$ ($\dot{m}\sim 0.4$).
The observed spectra are well reproduced with the model, and require
an outer standard disc as well as a warm Comptonisation
component. This is different to conclusions from previous spectral
fits as we constrain our warm Comptonisation component to have seed
photons and luminosity from an underlying disc rather than
allowing these to be free parameters. The midplane disc is generally
passive i.e. the seed photons are produced by reprocessing rather than
intrinsic dissipation, which sets the spectral index to
$\Gamma_{\rm warm}=2.5$ \citep{petrucci2017}. The transition between the
standard disc and warm Comptonisation is always at temperatures
consistent with the peak in UV opacity which might point to its origin
in the changing disc structure due to failed UV line driven winds
\citep{laor2014}. The hot plasma has almost constant dissipation,
consistent with 0.02--$0.04L_{\rm Edd}$ for all $\dot{m}$. This implies a
smaller size scale with increasing $\dot{m}$, as inferred from X-ray
variability \citep{mchardy2006}. The hard X-ray spectral index is
consistent with this dissipation always taking place in a region with
no underlying disc i.e. a truncated disc/hot inner flow geometry.

Fixing this derived geometry gives a full SED model which depends only
on mass and mass accretion rate. This model successfully explains the
observed tight UV/X relation shown by \cite{lusso2017} as a
combination of the constant hard X-ray dissipation together with
selection effects which mean that the rarer, higher mass quasars are
seen preferentially at larger distances so require higher Eddington
fractions to be detected. This selection effect introduces scatter and
bias but our model gives a physically based understanding of these
factors, so that the relation can be used to probe cosmology.  

The model includes the contribution to the optical/UV flux from 
X-ray illumination of the outer disc and warm Comptonisation
region. We calculate the optical variability resulting from a
stochastic change in X-ray flux. This predicts the fast variability in
optical should be a strongly decreasing function of Eddington
fraction, as the fixed (average) hard X-ray dissipation is a smaller
fraction of the bolometric luminosity at higher Eddington ratios. This
matches to some of the trends seen in systematic surveys of AGN
variability e.g., SDSS Stripe 82 \citep{macleod2010}, but there is more
variability seen in high Eddington fraction AGN than predicted. This
probably indicate that such highly radiation pressure dominated discs
are somewhat unstable. This should motivate theoretical studies to
give better understanding of such discs. 

\section*{Acknowledgements}


We thank to M. Mehdipour  for providing us SEDs in
\cite{509,5548} and helpful comments on the data.  We are also grateful to
C. Macleod and E. Lusso for providing us the data points in
\cite{macleod2010} and \cite{lusso2017}, and useful discussions and
comments.  Special thanks to H. Noda and C. Jin for helpful discussions.  AK is
supported by research program in foreign country by Shibaura Institute
of Technology. CD acknowledges support from STFC (ST/P000541/1), and
useful conversations with O. Blaes and J.M. Hameury. 
We also thank P.O.Petrucci as our referee for valuable comments.




\bibliographystyle{mnras}



\appendix
\section{Parameters of the Model}

We show all the spectral parameters of {\sc agnsed} in table~\ref{tab:appendix}. 
In the public version of {\sc agnsed}, albedo is fixed at $a=0.3$. 
Seed photon temperature for the hot Comptonisation component is calculated 
internally (see section \ref{sec:pl}).
$r_{\rm hot}$ is adopted as a fitting parameter instead of $L_{\rm diss, hot}$.
The model has some switching parameters. 
If  parameter 6 is negative, the model gives  the hot comptonisation component.
And if parameter 7 is negative, the model gives the warm comptonisation component.
If parameter 9 is negative, the model gives the outer disc.
If parameter 12 is negative, the code will use the self gravity radius as calculated from \cite{laor1989}.

{\sc qsosed} is the simplified version of {\sc agnsed} by fixing some parameters at the typical values and by including reprocessing.
The spectral parameters are shown in table~\ref{tab:appendix2}. 
The rest of the parameters are fixed at $kT_{e, {\rm hot}}=100$~keV,  $kT_{e, {\rm warm}}=0.2$~keV, $\Gamma_{\rm warm}=2.5$, $r_{\rm warm}=2r_{\rm hot}$, $r_{\rm out}=r_{\rm sg}$ and $h_{\rm max}=100$. 
Also, $\Gamma_{\rm hot}$ is calculated via eq.(\ref{eq:gamma}) and $r_{\rm hot}$ is calculated to satisfy $L_{\rm diss,hot}=0.02L_{\rm Edd}$. 

\begin{table}
	\centering
	\caption{Parameters in {\sc agnsed}. }
	\label{tab:appendix}
	\begin{tabular}{lp{7cm}} 
		\hline
par1 &mass, black hole mass in solar masses\\
par2 &dist, comoving (proper) distance in Mpc\\
par3 & $\log\dot{m}$, $\dot{m}=\dot{M}/\dot{M}_{\rm Edd}$ where $\eta\dot{M}_{\rm Edd}c^2=L_{\rm Edd}$\\
par4 & $a^\ast$, dimensionless black hole spin\\
par5 & $\cos i$, inclination angle $i$ for the warm Comptonising component and the outer disc.\\
par6 & $kT_{e,\rm hot}$, electron temperature for the hot Comptonisation component in keV.
If this parameter is negative then only the hot Comptonisation component is used. \\
par7 & $kT_{e,\rm warm}$, electron temperature for the warm Comptonisation component in keV.
If this parameter is negative then only the warm Comptonisation component is used. \\
par8 & $\Gamma_{\rm hot}$, spectral index of the hot Comptonisation component.
If this parameter is negative, the code will use the value calculated via eq.(\ref{eq:gamma}). \\
par9 & $\Gamma_{\rm warm}$, spectral index of the warm Comptonisation component.
If this parameter is negative then only the outer disc component is used. \\
par10 &  $r_{\rm hot}$, outer radius of the hot Comptonisation component in $R_{\rm g}$\\
par11 & $r_{\rm warm}$, outer radius of the warm Comptonisation component in $R_{\rm g}$\\
par12 & $\log r_{\rm out}$, $\log$ of the outer radius of the disc in units of $R_{\rm g}$. 
If this is parameter is negative, the code will use the self gravity radius as calculated from \cite{laor1989}.\\
par13 & $h_{\rm max}$, the upper limit of the scaleheight for the hot Comptonisation component in $R_{\rm g}$. If this parameter is smaller than parameter 10, the hot Comptonisation region is a sphere of radius $h_{\rm max}$ by keeping $L_{\rm diss, hot}$ determined by $r_{\rm hot}$ via eq.(\ref{eq:Lhotdiss}).\\
par14 &switching parameter for the reprocessing, 0 or 1. If this parameter is 0, reprocessing is not considered. If this parameter is 1, reprocessing is included.\\
par15 & Redshift\\
\hline
\end{tabular}
\end{table}

\begin{table}
	\centering
	\caption{Parameters in {\sc qsosed}. }
	\label{tab:appendix2}
	\begin{tabular}{lp{7cm}} 
		\hline
par1 &mass, black hole mass in solar masses\\
par2 &dist, comoving (proper) distance in Mpc\\
par3 & $\log\dot{m}$, $\dot{m}=\dot{M}/\dot{M}_{\rm Edd}$ where $\eta\dot{M}_{\rm Edd}c^2=L_{\rm Edd}$\\
par4 & $a^\ast$, dimensionless black hole spin\\
par5 & $\cos i$, inclination angle $i$ for the warm Comptonising component and the outer disc.\\
par6 & Redshift\\
\hline
\end{tabular}
\end{table}


\bsp	
\label{lastpage}
\end{document}